\documentclass[journal=jacsat,manuscript=article]{achemso}

\usepackage{chemformula}
\usepackage[T1]{fontenc}
\usepackage{amssymb}
\usepackage{graphicx}
\usepackage{times}
\usepackage{xcolor}
\usepackage{textcomp,gensymb}
\usepackage{multirow}

\author{\small Letizia Tavagnacco}
\affiliation{CNR-ISC, Uos Sapienza, Piazzale A. Moro 2, 00185 Roma, Italy}
\alsoaffiliation{Dipartimento di Fisica, Sapienza Università di Roma, Piazzale A. Moro 2, 00185 Roma, Italy}
\altaffiliation{These authors contributed equally to this work}
\author{Elena Buratti}
\affiliation{Department of Environmental and Prevention Sciences, University of Ferrara, Via L. Borsari, 46, 44121 Ferrara, Italy}
\altaffiliation{These authors contributed equally to this work}
\author{Jacopo Vialetto}
\affiliation{Dipartimento di Chimica "Ugo Schiff", Università di Firenze,  Sesto Fiorentino (FI), 50019 Italy}
\alsoaffiliation{Consorzio per lo Sviluppo dei Sistemi a Grande Interfase (CSGI), via della Lastruccia 3, Sesto Fiorentino (FI), 50019, Italy}
\author{Francesco Brasili}
\affiliation{CNR-ISC, Uos Sapienza, Piazzale A. Moro 2, 00185 Roma, Italy}
\alsoaffiliation{Dipartimento di Fisica, Sapienza Università di Roma, Piazzale A. Moro 2, 00185 Roma, Italy}
\author{Elisa Ballin}
\affiliation{CNR-ISC, Uos Sapienza, Piazzale A. Moro 2, 00185 Roma, Italy}
\alsoaffiliation{Dipartimento di Fisica, Sapienza Università di Roma, Piazzale A. Moro 2, 00185 Roma, Italy}
\author{Kuno Schw{\"a}rzer}
\affiliation{J\"{u}lich Centre for Neutron Science (JCNS), Forschungszentrum J\"{u}lich GmbH, Wilhelm-Johnen-Stra{\ss}e, 52428 J\"{u}lich, Germany}
\author{Jitendra Mata}
\affiliation{Australian Centre for Neutron Scattering (ACNS), Australian Nuclear Science and Technology Organisation (ANSTO), Lucas Height, New South Wales 2234, Australia}
\alsoaffiliation{School of Chemistry, University of New South Wales, NSW, Australia}
\author{Graziano Di Carmine}
\affiliation{Department of Environmental and Prevention Sciences, University of Ferrara, Via L. Borsari, 46, 44121 Ferrara, Italy}
\author{Monica Bertoldo}
\affiliation{Department of Chemical, Pharmaceutical and Agricultural Sciences, University of Ferrara, via L. Borsari 46, 44121 Ferrara, Italy}
\author{Ester Chiessi}
\affiliation{Department of Chemical Science and Technologies, University of Rome Tor Vergata, Via della Ricerca Scientifica I, 00133 Rome, Italy.}
\author{Marco Laurati}
\affiliation{Dipartimento di Chimica "Ugo Schiff", Università di Firenze,  Sesto Fiorentino (FI), 50019 Italy}
\alsoaffiliation{Consorzio per lo Sviluppo dei Sistemi a Grande Interfase (CSGI), via della Lastruccia 3, Sesto Fiorentino (FI), 50019, Italy}
\email{marco.laurati@unifi.it}
\author{Emanuela Zaccarelli}
\affiliation{CNR-ISC, Uos Sapienza, Piazzale A. Moro 2, 00185 Roma, Italy}
\alsoaffiliation{Dipartimento di Fisica, Sapienza Università di Roma, Piazzale A. Moro 2, 00185 Roma, Italy}
\email{emanuela.zaccarelli@cnr.it}

\title{\large
Thermoresponsive copolymer microgels synthesized via single-step precipitation polymerization: random or block structure?}

\begin{document}

\begin{abstract}
The inner structure of polymeric particles critically influences their phase behavior and functionality, governing their mechanical properties and their physical and chemical interactions. For thermoresponsive microgels, i.e. colloidal particles comprising a crosslinked polymer network that undergo a volume transition upon temperature changes, structural control is key to tailor the material responsivity and broaden the range of applications. In this work, we present a comprehensive investigation of the internal structure of poly(N-isopropylacrylamide -\textit{co}-N-isopropylmethacrylamide), P(NIPAM-\textit{co}-NIPMAM), copolymer microgels, combining small-angle neutron scattering (SANS), dynamic light scattering (DLS), and nuclear magnetic resonance (NMR) measurements with multi-scale simulations. By synthesizing different samples, including isotopically labeled microgels via selective deuteration, we probe the microgels swelling behavior and the evolution of their internal architecture as a function of temperature, revealing distinct signatures of the individual polymers. To elucidate their internal distribution, we perform monomer-resolved microgel simulations across different copolymer models. A direct comparison between experimental and numerical form factors under different, neutron-selective conditions provides evidence of a preferential organization into block structures rather than a random arrangement. These results are confirmed by $^{13}$C-NMR which reveals the clear presence of NIPAM blocks within a more random arrangement of the remaining monomers and by atomistic molecular dynamics simulations on copolymer chains, which also shed light on a possible origin in the dependence of the hydrogen bonding capability on the local environment.
These findings provide a detailed microscopic picture of the inner architecture of P(NIPAM-\textit{co}-NIPMAM) microgels, revealing an unexpected structural organization that may be generalized to other copolymer systems and could be promising to tailor microgel design and enhance control of material responsivity.
\end{abstract}

\section{Introduction} 
Thermoresponsive microgels represent a promising class of soft colloids, internally composed of cross-linked polymer networks. These systems function as smart polymeric materials, exhibiting a sharp and reversible volume change at the so-called volume phase transition temperature (T$_{VPT}$). The characteristic response to external stimuli arises from the lower critical solution temperature (LCST) phase behavior of the constituent polymer. The thermosensitivity of polymers in solution remains a central focus of research for the design of materials tailored to specific functions. Indeed, responsive polymeric materials are increasingly contributing to a broad range of applications, such as drug delivery, diagnostics, tissue engineering, as well as optical systems, and biosensors~\cite{stuart2010emerging,plamper2017functional,karg2019nanogels,wolter2025asymmetric}. Among the thermoresponsive polymers, poly(N-isopropylacrylamide) (PNIPAM) is the most intensively studied because its lower critical solution temperature in water is $\sim$305 K,~\cite{halperin2015poly} close to the physiological temperature.

The inner structure of polymeric particles plays a crucial role in determining their behavior and functionality, as it governs their mechanical response and as well as their physical and chemical interactions. It is well-established that the synthetic methods used in the preparation of microgels strongly influences the resulting internal architecture~\cite{meyer2005influence,kyrey2019inner}. Small angle neutron scattering experiments (SANS) have shown that the experimental procedure commonly used for synthesizing PNIPAM microgels, based on precipitation polymerization~\cite{pelton2000temperature}, yields particles with a non-uniform internal distribution characterized by a core–corona morphology~\cite{stieger2004small}. This structural heterogeneity arises from a significantly faster reaction kinetics of the cross-linker, typically N,N-methylenebisacrylamide (BIS), as compared to that of the N-Isopropylacrylamide (NIPAM) monomers. Recently, numerical simulations of \emph{in silico} microgels~\cite{gnan2017silico} have successfully reproduced the internal structure of PNIPAM microgels with quantitative accuracy~\cite{ninarello2019modeling,hazra2023structure}.

Incorporating different comonomers into microgels allows one to tune their morphology and the interactions occurring both within and among the particles. This, in turn, enables to tailor the material responsivity, broadening the range of possible applications. Several thermoresponsive copolymer microgels have been synthesized, providing the possibility to vary the volume phase transition temperature in a controlled way~\cite{hertle2013thermoresponsive,hannappel2021smart}. In particular, T$_{VPT}$ can also be adjusted by changing the comonomer content~\cite{inomata1995swelling,friesen2022modified}. A commonly used comonomer is N-Isopropylmethacrylamide (NIPMAM), which differs from NIPAM only for the presence of an additional methyl group in the $\alpha$ carbon of the carbonyl, as shown by the chemical structure reported in Figure~\ref{fig:Exp_FF}a. This structural modification results in a higher LCST of $\sim$317 K~\cite{ko2020temperature}, about 10 K above that of PNIPAM. However, the introduction of an additional monomer may  influence the internal structure of the resulting microgels.
A pioneering neutron scattering work on copolymer microgels composed by equimolar amounts of NIPAM and NIPMAM (P(NIPAM-\textit{co}-NIPMAM)) was put forward several years ago~\cite{JACSRich}, but the missing selectivity of the measurements with respect to the two polymers did not allow to discriminate their relative arrangements, particularly at low temperatures, where hydrophobic interactions do not play a major role. This study was unsurprisingly not followed by subsequent investigations of these specific microgels aimed to address the distribution of the constituent copolymers within the microgel. Instead, a Fourier-transform infrared (FTIR) spectroscopy study on copolymer microgels based on NIPMAM and NIPAM has demonstrated that the internal architecture of copolymer microgels has also a fundamental importance in determining their phase behavior\cite{wiehemeier2019swelling}. Specifically, it was found that while statistical copolymer microgels lead to a cooperative phase transition corresponding to a collapse at a specific temperature, some core-shell microgel particles exhibit a linear response of the hydrodynamic radius with temperature\cite{wiehemeier2019swelling}. Conversely, copolymer microgels based on N-vinylcaprolactam (VCL) and NIPAM or NIPMAM with various compositions, also probed by $^1$H transverse magnetization relaxation, were found to have a similar internal structure, characterized by fuzzy surfaces and dense cores, in which the two monomers were homogeneously present, independently of the chemical composition.~\cite{balaceanu2013copolymer,wu2014behavior}
Furthermore, while coarse-grained and atomistic simulations have both been extensively applied to investigate the phase behavior of hopolymer PNIPAM and PNIPMAM~\cite{ninarello2019modeling,tavagnacco2025understanding}, numerical studies specifically targeting copolymer systems are still lacking.

The present work addresses these long-standing questions, aiming to provide a fundamental understanding of the internal architecture of thermoresponsive P(NIPAM-\textit{co}-NIPMAM) microgels. To this end, we perform a combined multi-scale experimental and numerical investigation, focusing on microgels at equimolar composition. By synthesizing different samples, including isotopically substituted ones by selectively deuterating one of the two constituent repeating units, we carry out Dynamic Light Scattering (DLS), Nuclear Magnetic Resonance (NMR) and small-angle neutron scattering (SANS) experiments. In this way we are able to probe the overall swelling behavior of the microgels as well as the evolution of their internal architecture as a function of temperature, revealing important signatures of the individual polymers. To shed light on their internal distribution, we then perform monomer-resolved microgel simulations across different copolymer models. The direct comparison between experimental and numerical form factors under the different, neutron-selective conditions allows us to identify a preferential distribution of P(NIPAM-\textit{co}-NIPMAM) microgels into block structures, as compared to a random arrangement. To validate these findings we then perform $^{13}$C-NMR experiments, which unambiguously reveal the presence of a large amount of NIPAM blocks, while NIPMAM is organized within a more random environment. Finally, we resort to atomistic molecular dynamics simulations of copolymer chains, which also support preferential block organization and highlight a correlation between hydrogen bonding capability of the constituent polymers in the copolymer environment.
Overall, our findings offer a detailed microscopic picture of the internal architecture of thermoresponsive copolymer microgels, showing evidence of an unexpected scenario in their relative arrangement, which could be generic for many different kinds of copolymer microgels.  Given the wide interest on these systems for various purposes~\cite{hertle2013thermoresponsive,forg2022copolymerization,ruiz2023concentration,kruger2024volume}, it is mandatory to take into account the possibility of an altered structure with respect to a fully random picture, in order to control and optimize the responsivity and usage of the materials.

\section{Results and discussion}

\subsection{Preparation and characterization of P(NIPAM-\textit{co}-NIPMAM) microgels}
\begin{figure*}
\centering
\includegraphics[width=1\linewidth]{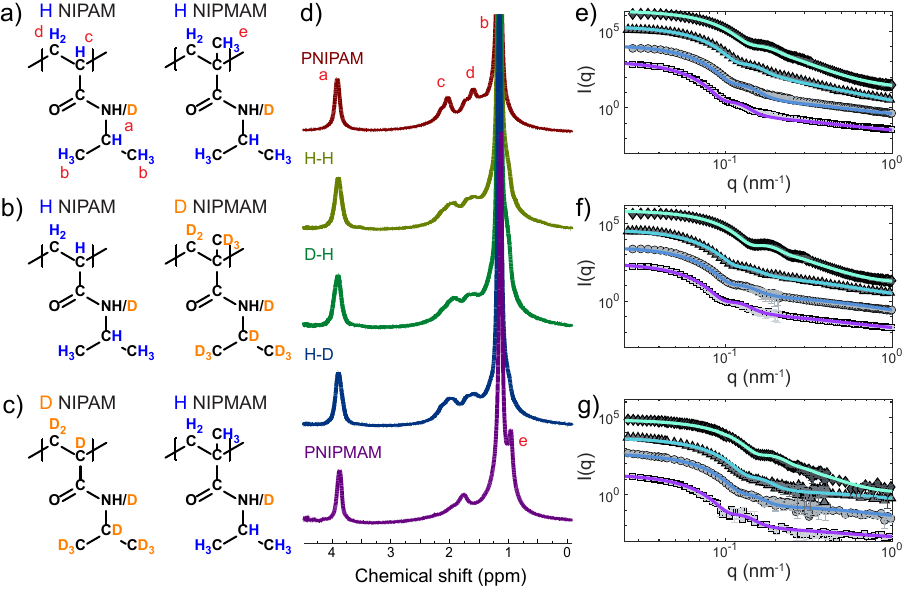}
\caption{a-c) Chemical formula of the repeating units of the polymers obtained from the indicated monomers. d) $^1$H-NMR spectra at 25°C of microgels dispersions in D$_2$O. Red letters indicate peak assignment with respect to the repeating units in a). e-g) SANS scattering profiles in D$_2$O as a function of temperature for P(H-NIPAM-\textit{co}-H-NIPMAM), H-H (e), P(H-NIPAM-\textit{co}-D-NIPMAM), H-D (f) and P(D-NIPAM-\textit{co}-H-NIPMAM), D-H (g). Lines are fits to the fuzzy sphere + blob scattering model of Eq. \ref{eq:Pfuzzy}.  Experiments are performed at 20°C (light gray squares, purple lines), 35°C (gray circles, blue lines), 37.5°C (dark gray triangles, light blue lines), and 50°C (black diamonds, green lines). Data at different T values are vertically shifted for clarity.}
\label{fig:Exp_FF}
\end{figure*}

Equimolar copolymer microgels with different isotopic (hydrogen/deuterium) compositions were synthesised in order to distinguish the individual contribution of NIPAM and NIPMAM using neutron scattering experiments. Specifically, microgels were prepared at constant crosslinker concentration ($c=5~mol~\%$) by copolymerizing (i) hydrogenated NIPAM and NIPMAM (P(H-NIPAM-\textit{co}-H-NIPMAM), H-H), (ii) hydrogenated NIPAM and deuterated NIPMAM (P(H-NIPAM-\textit{co}-D-NIPMAM), H-D), and (iii) deuterated NIPAM and hydrogenated NIPMAM (P(D-NIPAM-\textit{co}-H-NIPMAM), D-H). The corresponding chemical structures are shown in Figures~\ref{fig:Exp_FF}(a-c). Details of the microgel synthesis and of the preparation of deuterated monomers are reported in section "Models and Methods" and in the SI.
DLS confirmed the thermoresponsivity of the microgels (see Figure S1a), which present a single transition temperature in H$_2$O at 38.4 $\pm$ 0.1$\degree$C for H-H, 39.1 $\pm$ 0.1$\degree$C for H-D and 39.9 $\pm$ 0.1$\degree$C for D-H, as estimated by fitting the deswelling curves with a sigmoidal function. The T$_{VPT}$ of the H-H microgels  approximately corresponds to the average of the transition temperatures of the homopolymer microgels~\cite{buratti2022role,cors2019deuteration}.

$^1$H-NMR (Figure~\ref{fig:Exp_FF}d and Figure S2) was used to quantify the molar fraction of the two components in the H-H copolymer microgel by using eq. \ref{eq:NMRratio} (see details in Materials and Methods). We obtain a NIPAM fraction of 49\% and a corresponding NIPMAM fraction of 51\%, in good agreement with the targeted 50\%-50\% composition. The $^1$H-NMR spectra of D-H and H-D copolymer microgels further indicate that when using deuterated monomers a comparable composition is obtained in the resulting microgels.

SANS experiments were performed on dilute suspensions of P(NIPAM-\textit{co}-NIPMAM) microgels in D$_2$O.
Inclusion of deuterated NIPAM and NIPMAM allowed us to perform contrast variation and investigate the copolymer microstructure within the microgel. Indeed, while fully hydrogenated copolymer microgels P(H-NIPAM-\textit{co}-H-NIPMAM) present a full contrast for both components, in partially deuterated copolymer microgels (P(H-NIPAM-\textit{co}-D-NIPMAM) and P(D-NIPAM-\textit{co}-H-NIPMAM)) the deuterated species is nearly contrast-matched in D$_2$O. Figures~\ref{fig:Exp_FF}(e-g) display the experimental SANS intensity profiles measured for the copolymer microgel suspensions at four characteristic temperatures corresponding to different phase states: the swollen state (20$\degree$C), the transition region (35$\degree$C and 37.5$\degree$C), and the collapsed state (50$\degree$C). For all copolymer microgels, intensities for T = 20$\degree$C show a $q$-dependence that is characterized by an initial decrease leading to a minimum, followed by a maximum and a power-law dependence at the largest $q$-values. This $q$-dependence is similar to that observed previously for homopolymeric PNIPAM microgels with a core-corona structure, typically described using a fuzzy sphere model~\cite{stieger2004small}. A shift of the form factor minima to higher $q$ values with increasing temperature, and a sharpening of the following maxima is observed, revealing progressive particle shrinkage and compaction. The experimental SANS intensities of dilute suspensions can be expressed as:

\begin{equation}
    I(q)= \phi V(\Delta\rho)^2P(q)
\label{eq:Iq}
\end{equation}

where $\phi$ is the particle volume fraction, V is the volume and $\Delta\rho = \rho_{\mu}-\rho_{s}$ is the contrast between the scattering length density of the microgels ($\rho_{\mu}$) and the solvent ($\rho_s$). The particle form factor was modeled using the fuzzy sphere function of Eq.\ref{eq:Pfuzzy}, with the fits shown as solid lines in Figures~\ref{fig:Exp_FF}(e-g). The fitting results are in good agreement with the experimental data over the entire explored $q$-range, indicating that all copolymer microgels have fuzzy surfaces and dense cores. The fitted parameters are listed in Tab. S2 of the SI.

As expected from the qualitative discussion of the SANS intensities, the particle radius and the fuzziness decrease for all samples with increasing T. We notice that very similar fit parameters are obtained for the fully hydrogenated and for the partially deuterated microgels, supporting the closely comparable internal structure of the three samples. In addition, the mesh size estimated from the polymer scattering term, $\zeta$, seems to display maximum values close to the VPT temperature, in analogy with other
copolymer microgels~\cite{hyatt2015segregation}.

\subsection{P(NIPAM-\textit{co}-NIPMAM) microgels from monomer-resolved simulations}
Having established the global core-corona structure of all synthesised P(NIPAM-\textit{co}-NIPMAM) microgels, we now turn our attention to investigate the inner topology of the polymer network, in order to understand the local distribution of each copolymer. To this aim, we exploit monomer-resolved simulations of copolymer microgels having the same copolymer ratio of 50\% as in the experiments. This coarse-grained model also accounts for the presence of a disordered network with the experimentally observed core–corona distribution.

\begin{figure*}
\centering
\includegraphics[width=1\linewidth]{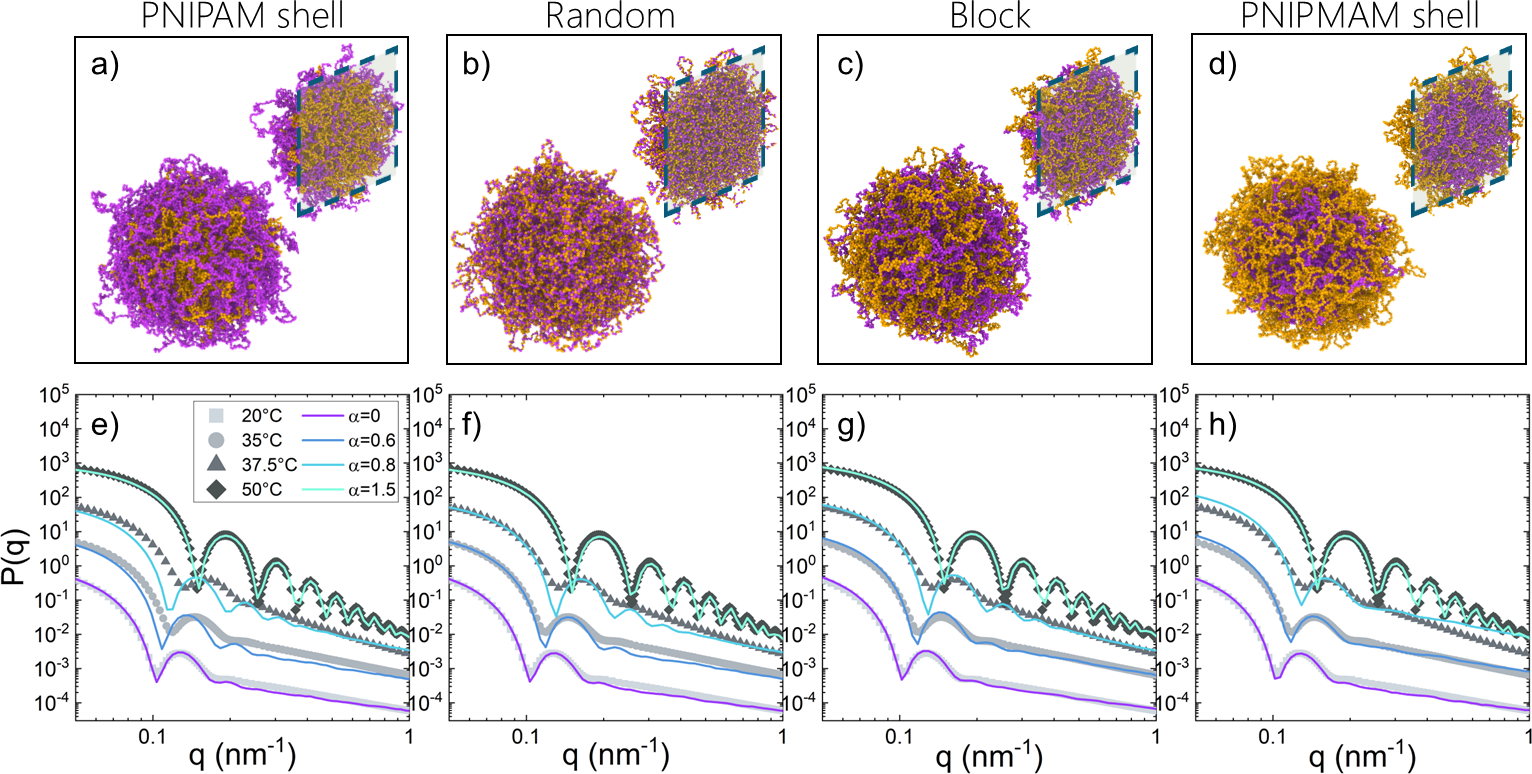}
\caption{Representative snapshots from the monomer-resolved simulations showing the different microgels architectures: (a) PNIPAM shell, (b) random, (c) block, and (d) PNIPMAM shell. PNIPAM and PNIPMAM particles are shown in purple and yellow, respectively. Cross-linkers are shown in purple. The inner part of the microgels structures is also highlighted with a sliced representation on the top panels.
e - f) Comparison between experimental and numerical form factors (P(q)) calculated for the four different structures. Experimental data are measured for fully protiated polymer samples (H-H) at 20°C (light gray squares), 35°C (gray circles), 37.5°C (dark gray triangles), and 50°C (black diamonds) and compared to numerical data calculated at $\alpha$ = 0.0 (purple lines), $\alpha$ = 0.6 (blue lines), $\alpha$ = 0.8 (light blue lines), and $\alpha$ = 1.5 (cyan lines). Data at different T values are vertically shifted for clarity. The experimental data are shown after subtracting smearing and polydispersity from the fuzzy sphere model fits.
}
\label{fig:models-ff}
\end{figure*}

We focus on four distinct network topologies of P(NIPAM-\textit{co}-NIPMAM) microgels, each representing a limiting case scenario for the internal copolymer organization, illustrated in Figures \ref{fig:models-ff}(a-d). Specifically, we consider: (i) a random network, where NIPAM and NIPMAM repeating units are randomly distributed throughout the network, named hereafter {\it random} (Figure \ref{fig:models-ff}b); (ii) a topology where each polymer chain included between two cross-linkers is entirely assigned to a PNIPAM or PNIPMAM homopolymeric segment, called {\it block} topology (Figure \ref{fig:models-ff}c); (iii) a {\it PNIPAM shell} topology, for which NIPAM repeating units are mostly distributed in the outer microgels corona (Figure \ref{fig:models-ff}a) and (iv) a {\it PNIPMAM shell} one where the opposite structure is employed, with NIPMAM repeating units all found in the outer corona (Figure \ref{fig:models-ff}d). The latter two cases (a and d) would be realized if a demixing between the two polymer blocks happens within the experimental synthesis, due to a marked difference in reaction rates or to mutual interactions. If this was the case, one would intuitively expect that the slightly smaller NIPAM units, having also a faster reaction kinetics~\cite{duracher1999preparation}, would go toward the center of the network in a PNIPMAM shell arrangement. We report also the opposite scenario for a full comparison of all possibilities.
On the other hand, the former two options (b and c) would correspond to a full mixing of the monomers, which would bind to each other either with the same chemical affinity (random case) or with a strong preference for the like-monomer (block case). Hence, the latter situation would correspond to the preferential formation of homopolymer chains, which are then joined by crosslinkers in the microgel network. By constructing these extreme copolymer models, we aim to enhance their distinctive structural signatures, thereby enabling a clearer discrimination among possible topologies within the finite size resolution of the coarse-grained models.  While it is well-established that core-shell microgels containing PNIPAM and PNIPMAM are synthesized via a two-step, seed and feed, polymerization process~\cite{berndt2005structure,berndt2006temperature}, the limiting cases presented here allow us to identify and characterize preferential organizational trends that could emerge even in a single-step copolymerization.

We first monitor the swelling behavior of the four P(NIPAM-\textit{co}-NIPMAM) microgels topologies, systematically varying the solvophobic parameter $\alpha$, which serves as an effective proxy for temperature (see Methods and SI for further details). To this aim, we calculate the hydrodynamic radius using a numerical procedure which has been previously validated~\cite{del2021two}, in analogy to the DLS measurements. As shown in Figure S1b, all different topologies display a sigmoidal dependence of the hydrodynamic radius on the solvophobic parameter $\alpha$.  We find that the VPT occurs within a range of $\alpha$ values that lie between those observed for PNIPAM and PNIPMAM homopolymer microgels.
Interestingly, in the case of shell arrangement of either of the two polymers, the copolymer in the shell region is found to strongly influence the overall behavior, shifting the occurrence of the VPT towards the corresponding homopolymer system. Conversely, the {\it random} and {\it block} microstructures are characterized by swelling curves that are intermediate between those of homopolymer PNIPAM and PNIPMAM microgels, in qualitative agreement with the DLS experimental behavior, but virtually indistinguishable from each other.
Hence, on the basis of these results, it is impossible to discriminate which internal structure would best describe the experimental samples.

\subsection{Inner structure of P(NIPAM-\textit{co}-NIPMAM) microgels from form factors}
To make a decisive step forward into the knowledge of the internal architecture of P(NIPAM-\textit{co}-NIPMAM) microgels, we combine monomer-resolved simulations with SANS experiments. Figures~\ref{fig:models-ff}(e-h) report a direct comparison between the experimental form factors measured for P(H-NIPAM-\textit{co}-H-NIPMAM) microgels, where the smearing and the polydispersity have been subtracted, and the numerical ones computed for the four employed microgel topologies at the same four representative temperatures as in Figure \ref{fig:Exp_FF}. These are compared to the same values of the solvophobic parameter $\alpha$ for all numerical models (see Figure S21b) by aligning the position of the first peak of the numerical form factor with that of the experimental one. This procedure yields a unique scaling factor, that allows us to convert numerical units into real ones and that is maintained for all temperatures, in analogy with previous works~\cite{ninarello2019modeling}.

The comparison of the $P(q)$ for P(H-NIPAM-\textit{co}-H-NIPMAM) microgels, reported in Figures~\ref{fig:models-ff}(e-h) shows that all four model topologies reproduce the experimental data with an overall good level of agreement, particularly for the lowest and the highest studied temperatures.
At intermediate temperatures, some discrepancies emerge, especially for the \emph{PNIPAM shell} (Figure~\ref{fig:models-ff}e) and \emph{PNIPMAM shell} (Figure~\ref{fig:models-ff}h) topologies. However, the presence of some sharper and deeper peaks in the numerical $P(q)$ may be attributable to a weak polydispersity of the experimental microgels, which is not accounted in the \emph{in silico} description. Overall, the comparison suggests that the \emph{block} and \emph{random} topologies are both able to provide an accurate description of the experimental SANS form factors for P(H-NIPAM-\textit{co}-H-NIPMAM) microgels. Again, these non-selective measurements cannot distinguish between the two arrangements.

\begin{figure*}
\centering
\includegraphics[width=1\linewidth]{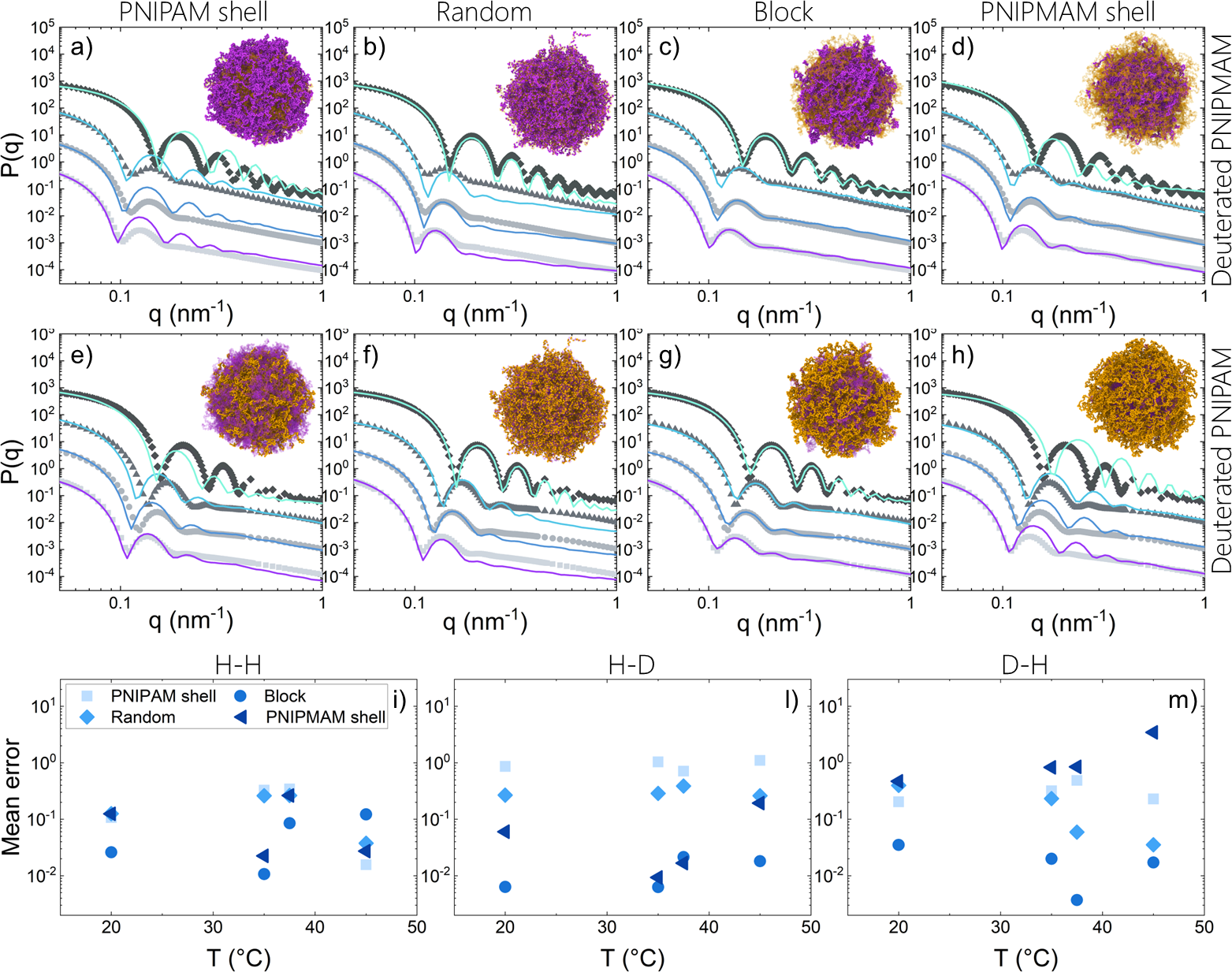}
\caption{Comparison between experimental and numerical form factors (P(q)).  Experimental data are measured for P(H-NIPAM-\textit{co}-D-NIPMAM) microgels (top row) at 20°C (light gray squares), 35°C (gray circles), 37.5°C (dark gray triangles), and 50°C (black diamonds) and
and for P(D-NIPAM-\textit{co}-H-NIPMAM) microgels  (second row) at at 20°C (light gray squares), 35°C (gray circles), 37.5°C (dark gray triangles), and 45°C (black diamonds). The experimental data are shown after subtracting smearing and polydispersity from the fuzzy sphere model fits and are
compared to numerical data for the four different models: (a,e) PNIPAM shell, (b,f) random, (c,g) block, and (d,h) PNIPMAM shell, calculated at $\alpha$ = 0.0 (purple lines), $\alpha$ = 0.6 (blue lines), $\alpha$ = 0.8 (light blue lines), and $\alpha$ = 1.5 (cyan lines).  Data at different T values are vertically shifted for clarity. The inset of each panel includes a representative snapshot from the monomer-resolved simulations calculated at $\alpha$ = 0.7, a condition representing the transition region. PNIPAM and PNIPMAM particles are shown in purple and yellow, respectively. Cross-linkers are shown in purple; (i-m) Estimated mean errors between experimental and numerical form factors in q range 0.05 - 0.5 nm$^{-1}$.}
\label{fig:ffnew}
\end{figure*}

This limitation is overcome by focusing on partially deuterated microgels.
Starting with P(H-NIPAM-\textit{co}-D-NIPMAM) microgels, for which the PNIPAM component is predominantly visible, the SANS $P(q)$, again after removeing resolution and polydispersity effects, are directly compared to simulations in Figure \ref{fig:ffnew} (top row). These plots  again confirm that both shell structures (Figure \ref{fig:ffnew}a for \emph{PNIPAM shell} and Figure \ref{fig:ffnew}d for \emph{PNIPMAM shell}, respectively) clearly fail to reproduce the experimental behavior, not being able to capture even the correct shift of the position of the first peak of the form factors. In particular, for the \emph{PNIPAM shell} topology, the predominant localization of PNIPAM in the outer shell results in a more collapsed structure, reflected by the shift of the form factor peaks to higher $q$. Conversely, for the \emph{PNIPMAM shell} topology, the swelling behavior driven by the PNIPMAM-rich shell yields peaks at lower values, corresponding to an overall more swollen structure than that observed experimentally. Instead, both the random (Figure \ref{fig:ffnew}b) and block (Figure \ref{fig:ffnew}c) topologies are able to correctly capture the relative positions of the peaks upon deswelling. The differences between these two models with respect to experiments are more subtle and amount to a better description of the overall peaks shape and large $q$-decay. For both features, the block model seems to perform better than the random one in comparison to experiments.

We complement these observations by also monitoring the SANS form factors measured for P(D-NIPAM-\textit{co}-H-NIPMAM) microgels, which primarily characterize the behavior of the PNIPMAM component, that are reported Figure~\ref{fig:ffnew} (second row). Again, we find that the experimental $P(q)$ strongly deviate from the simulated shell models, with a notable shift of the peaks positions to opposite directions as for P(H-NIPAM-\textit{co}-D-NIPMAM) microgels. The results for random and block topologies again confirm the previous observations. Indeed,
while the \emph{random} topology (Figure~\ref{fig:ffnew}f) still captures the qualitative trend of the SANS $P(q)$ and matches the shift of the first peak position with increasing temperature  at low $q$ value, it fails to accurately reproduce the features at intermediate and high $q$. In contrast, again the \emph{block} topology shows an overall semi-quantitative agreement with the experimental form factor across the full $q$ range.
Given that the measured form factors pertain to three different experimental samples and the simulations are performed with a unique model able to describe all three microgels, we believe that the present findings are robust against statistical error and strongly indicate that the block topology is the most accurate to reproduce the experimental behavior of P(NIPAM-\textit{co}-NIPMAM) microgels. These results point to the fact that there is a significant presence of local domain structuring, where identical polymer units (either NIPAM or NIPMAM) preferentially bind in sequence within the copolymer network, rather than being homogeneously distributed as in a fully random structure.
To provide a quantitative evaluation of the accuracy of the different models, we plot the estimated mean errors between experimental and numerical form factors in Figure \ref{fig:ffnew}(i-m). We find that, except for the high-$T$ H-H case, the error is minimized by the block configuration, which overall performs significantly better than the random case.

It is important to note that the \emph{block} topology model used for this analysis was developed to represent an extreme case of the microgel internal architecture, wherein the monomers are locally organized into domain structures. In particular, the choice to define domains consisting of polymer segments located between two cross-linkers was made somewhat arbitrarily, with the aim of generating sufficiently large regions. This approach is  necessary considering the smaller size of the numerical microgels as compared to their experimental counterparts. To further explore the role of domain size, we designed an additional topology, referred to as \emph{small block}, characterized by blocks containing roughly ten consecutive monomers of the same type, randomly distributed throughout the microgel network. The comparison between experimental and simulated form factors, reported in Figure S19, shows that the \emph{small block} topology also provides a satisfactory description of the experimental SANS data. Of course, it is to keep in mind the coarse-grained nature of the employed models: here each bead is a group of monomers with size roughly equal to the Kuhn length of the polymer. Hence to assign a block of several monomers does not necessarily lead to a 1:1 correspondence with the atomic scale but rather to an arrangement where a majority of the two species is present within the bead. These findings strengthen the hypothesis that preferential formation of homopolymer domains occurs within the microgel network. However, the analysis of the form factor alone does not allow for an unambiguous determination of the precise domain size. Such heterogeneities in the polymer network could arise from differences in the homopolymerization and copolymerization propagation kinetics of the two monomers~\cite{coote2002copolymerization}. In fact, monomers with comparable reactivity are inserted equally into propagating chains resulting with  random sequences. Instead, when monomers have different reactivity, the more reactive monomer is inserted preferentially, resulting in long sequence, up to its concentration in the reaction medium decrease. For the PNIPAM and PNIPMAM system, it has been shown that NIPAM exhibits a higher propagation rate than NIPMAM\cite{duracher1999preparation}, which could promote the formation of such heterogeneities in the resulting microgel network.

\begin{figure*}
\centering
\includegraphics[width=1\linewidth]{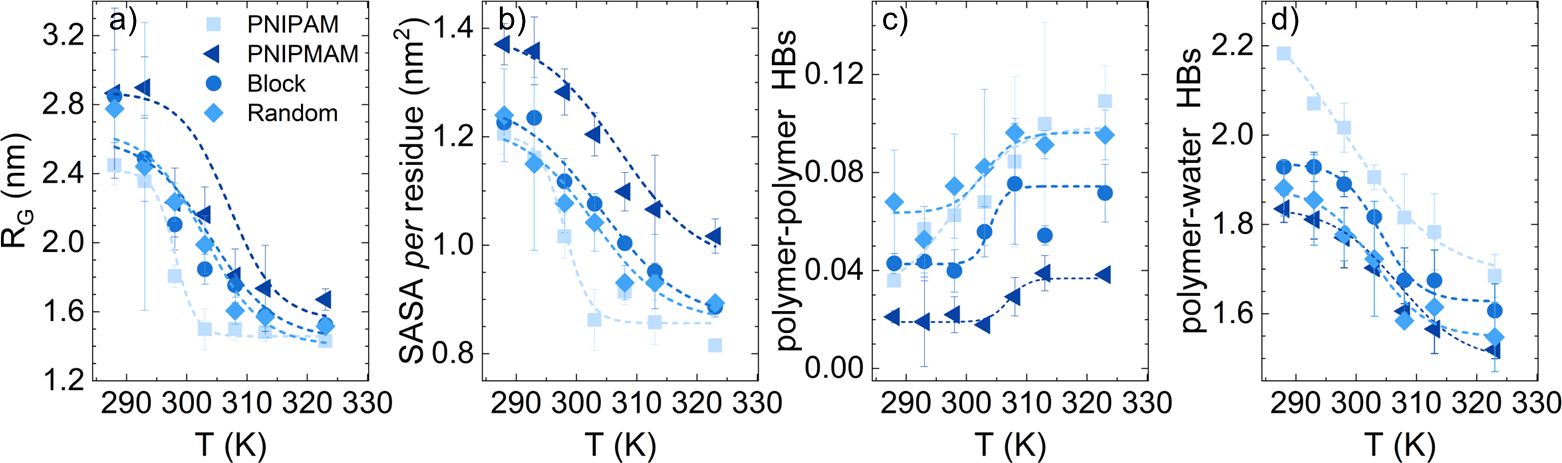}
\caption{Temperature dependence of (a) radius of gyration; (b) solvent accessible surface area; (c) number of polymer-polymer hydrogen bonds; and (d) number of polymer-water hydrogen bonds calculated from atomistic molecular dynamics simulations for a polymer chain composed by 90 repeating units of PNIPAM (gray squares), PNIPMAM (dark blue triangles) and a copolymer with a composition 1:1 and a random (light blue diamonds) or block (blue circles) microstructure. Data are averaged over two independent replicas.}
\label{fig:atomistic}
\end{figure*}

\subsection{Molecular insights from P(NIPAM-\textit{co}-NIPMAM) atomistic chains}
We now use atomistic molecular dynamics simulations to look for a molecular rationale for the preferential formation of domain structures within the microgel network. To this aim, we investigate the solution behavior of polymer chains composed by 90 repeating units. We employ a three-times larger system as compared to the conventional model of 30 repeating units~\cite{rovigatti2019numerical} and average results over two independent replicas to probe a representative ensemble of configurations. We examine the behavior of a
\emph{random} and \emph{block} P(NIPAM-\textit{co}-NIPMAM) chain with a comonomer molar ratio of 1:1, as well as homopolymer PNIPAM and PNIPMAM chains. In particular, we design the \emph{block} copolymer chain with two regions of each polymer composed by 45 repeating units. To characterize the coil-to globule transition, we monitor the radius of gyration and the solvent accessible surface area (SASA) as a function of temperature. As reported in Figures \ref{fig:atomistic}a and \ref{fig:atomistic}b, all chains display a reduction of size and SASA induced by temperature, with a sigmoidal dependence. The temperature range in which extended conformations are predominantly populated is larger for PNIPMAM, as compared to PNIPAM, while the \emph{random} and \emph{block} chains exhibit an intermediate behavior in between the two pure polymers. In all systems, the size of both globular and coil states increases with the addition of PNIPMAM.
To probe the molecular interactions driving these transitions, we also report the temperature dependence of the number of hydrogen bonds (HBs), i.e. polymer–polymer and polymer–water interactions in Figures \ref{fig:atomistic}c and \ref{fig:atomistic}d, that are also described by a sigmoidal function. From the global sigmoidal fit of all the observables, we obtain an estimate of the critical temperature, with $T_c$ values of 298$\pm$2 K, 307$\pm$3 K, 304$\pm$2 K, and 303$\pm$1 K, for pure PNIPAM, pure PNIPMAM, for the \emph{block} and for the \emph{random} chains, respectively. These transition temperatures are consistent with the experimental findings, with the critical temperature of the copolymers being intermediate to those of PNIPAM and PNIPMAM, as detected for the $T_{VPT}$ of copolymer microgels. Figures~\ref{fig:atomistic}c and \ref{fig:atomistic}d also reveal that PNIPAM and PNIPMAM have, respectively, the highest and lowest tendency to form both polymer-polymer and polymer-water HBs. Moreover, the \emph{block} copolymer again behaves in an intermediate fashion, as found for the structural observables. The intermediate behavior found for polymer-water HBs is consistent with the characterization of the interactions occurring between water and the polymer carried out by analyzing the N-H bond using FTIR measurements~\cite{friesen2022modified}. However, it is striking to observe that the \emph{random} chain does not share the same ``intermediate'' behaviour, but rather it is characterized by a number of polymer-polymer HBs much closer to that of pure PNIPAM and by a number of polymer-water HBs more similar to pure PNIPMAM.
These findings suggest that, at the molecular level, a random distribution of monomers does not adequately capture the experimental behavior. Instead, the formation of contiguous blocks of the same monomers appears to be a more realistic representation of the chains forming the microgel network.

\begin{figure*}
\centering
\includegraphics[width=1\linewidth]{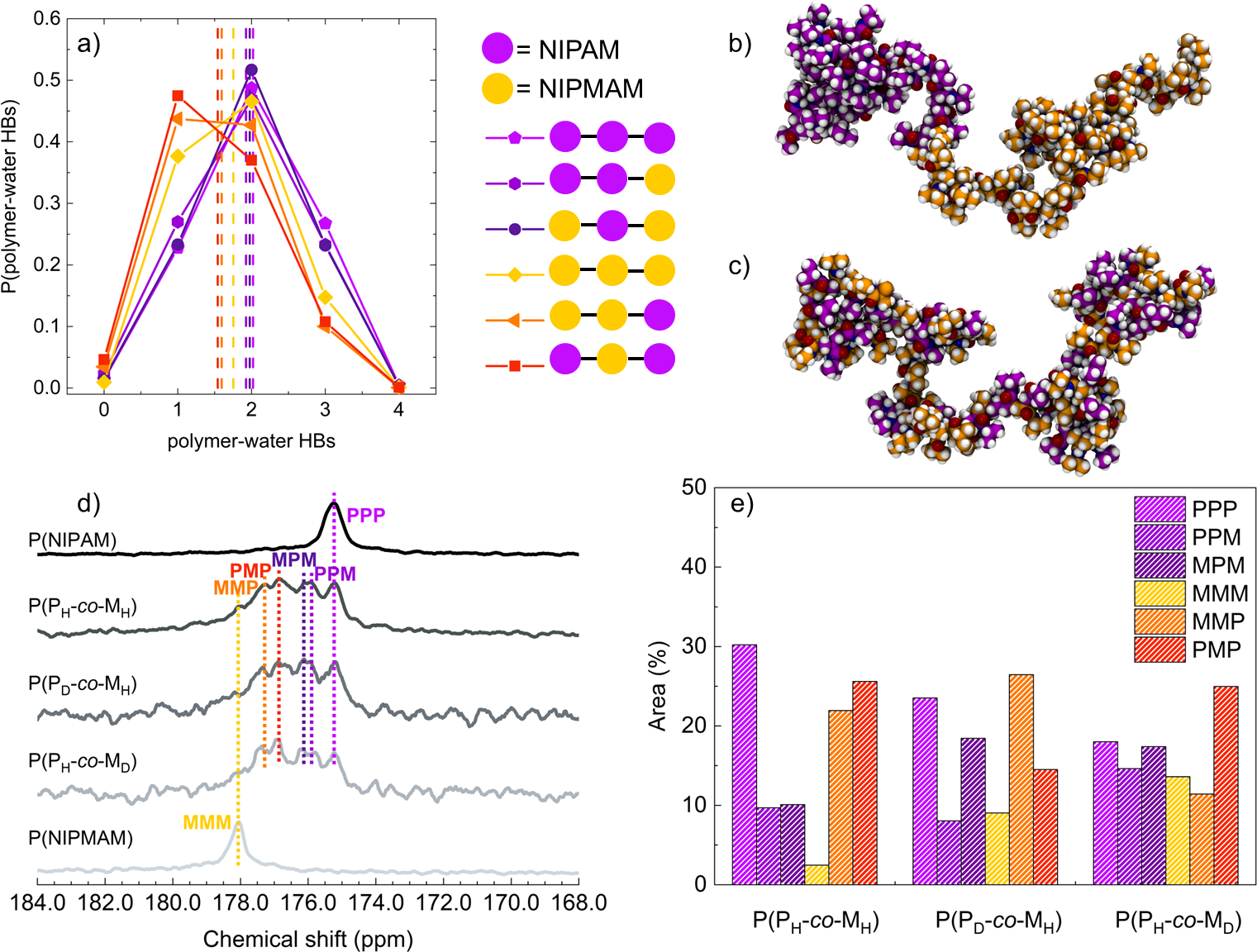}
\caption{(a) Probability distribution of the number of polymer-water hydrogen bonds formed by each repeating unit in the atomistic chains as a function of the neighbouring repeating units: (PPP) a NIPAM unit in between two NIPAM units (purple pentagons), (MMM) a NIPMAM unit in between two NIPMAM units (yellow diamonds); (MPM) a NIPAM unit in between two NIPMAM units (dark violet circles); (MPP) a NIPAM unit in between a NIPAM and a NIPMAM unit (violet hexagons); (MMP) a NIPMAM unit in between a NIPAM and a NIPMAM unit (orange triangles); and (PMP) a NIPMAM unit in between two NIPAM units (red squares). Data are calculated at 298 K and averaged over all model systems and for the two replicas. Vertical lines are averaged distribution values. Representative snapshots from atomistic simulations showing (b) a \emph{block} and (c) a \emph{random} P(NIPAM-\textit{co}-NIPMAM) chain in the coil state. Oxygen, nitrogen, hydrogen, NIPAM carbon, and NIPMAM carbon atoms are shown in red, blue, white, purple, and yellow, respectively. (d) Signals of carbonyl peak in $^{13}$C-NMR spectra of PNIPAM, P(H-NIPAM-\textit{co}-H-NIPMAM), P(D-NIPAM-\textit{co}-H-NIPMAM), P(H-NIPAM-\textit{co}-D-NIPMAM), PNIPMAM microgels from top to bottom, respectively, with peaks assignment due to the different sequences, PPP, PPM, MPM, PMP, MMP, MMM, from right to left. (e) Percentage distributions of the different sequences, obtained from the deconvolution of the carbonyl signal in the three copolymers.}
\label{fig:triad}
\end{figure*}

In order to support this rather unexpected result, we further ask ourselves
why such a random arrangement does not show an intermediate behavior, in line with expectations and experiments, whilst the block arrangement does.
To this aim, we examine how the local environment affects the probability to form hydrogen bonds of individual repeating units.
We focus on polymer-water HBs, as they are significantly more numerous than polymer–polymer HB interactions, thereby providing better statistical accuracy.  Figure~\ref{fig:triad}a shows the probability distribution of polymer–water hydrogen bonds for each NIPAM (denoted as P) and NIPMAM (denoted as M) repeating unit, classified on the basis of their neighboring repeating units and then averaged over all chain types. We then follow HBs formed by a NIPAM unit in a PPP, MPM and PPM environment, as well as those formed by a NIPMAM unit in a MMM, PMM and PMP configuration. The results clearly show that when NIPAM is linked to two other NIPAM units (PPP), the highest average number of polymer-water HBs is formed ($\sim$2.0) and also the largest probability of finding 3 HBs per residue is detected. Instead, when one (PPM) or two (MPM) adjacent units are substituted with NIPMAM, a weak decrease in the average number of polymer-water HBs is detected (down to $\sim$1.93). In contrast, a NIPMAM unit linked to two NIPMAM units (MMM) forms a significantly lower number of polymer-water HBs ($\sim$1.75). Strikingly, this tendency is enhanced by the presence of linked NIPAM residues (PMM and PMP), which strongly affects NIPMAM behavior, further reducing its hydrogen bonding capability down to $\sim$1.54.
Hence, at odds with expectations where surrounding NIPMAM with NIPAM would improve its affinity to water for a possible cooperative effect on water-polymer hydrogen bonding, this appears to be not the case and we actually find that a continuous alternation of NIPMAM with NIPAM units further inhibits its H-bonding ability with respect to water.

The clear dependence of the hydrogen bonding capability on the local monomeric sequence highlights the importance of the copolymer configuration. Due to the different response on the local environment, the \emph{block} copolymer chain (Figure~\ref{fig:triad}b) is thus able to form a number of polymer-water hydrogen bonds which is intermediate between the two pure polymers. Differently, the \emph{random} copolymer chain, due to frequent alternation of monomer types (Figure~\ref{fig:triad}c), is subjected to a significant reduction of polymer-water interactions leading to a behavior more similar to that of PNIPMAM homopolymer. These findings suggest that the interplay between local monomer composition and hydrogen-bonding behavior plays a crucial role in the hydration properties of copolymer chains and may influence the formation of contiguous domains of the same monomer type at the molecular level.

\subsection{Confirmation of partial block structure by $^{13}$C-NMR measurements}
To validate the findings obtained from atomistic molecular dynamics simulations and to investigate the possible sequences within the microgel network, detailed $^{13}$C-NMR measurements were carried out. As expected, the spectra of the copolymers microgels (Figures S9-11) show more signals than the simple superposition of those of the two homopolymers (Figures S7-8). This is due to the fact that the repeating units of NIPAM and NIPMAM can be linked either to identical units or to those of the other monomer. For a detailed analysis of the copolymer structure, the carbonyl signal between 175 and 178 ppm was selected, being the cleanest and most suitable one for a deconvolution process.
As a first step, it was verified that the carbonyl signals of NIPAM and NIPMAM microgels exhibit comparable intensity, in order to ensure their suitability for a quantitative estimation of the copolymer composition. To this end, a $^{13}$C spectrum was recorded for a mixture of the two homopolymer microgels, PNIPAM and PNIPMAM, with a known molar composition of 60\% PNIPAM and 40\% PNIPMAM (Figure S12). The deconvolution of the two carbonyl peaks (Figure S13) yielded an area ratio of 60\% for PNIPAM (signal at 175.29 ppm) and 40\% for PNIPMAM (signal at 178.06 ppm), thus confirming the composition of the test mixture and supporting the use of the carbonyl signals for further quantitative analysis.
In all the copolymer microgels (Figure~\ref{fig:triad} and Figure S9-S11), it can be observed that the carbonyl peak is actually composed of six distinct signals. The absence of two separate peaks, as seen in the spectrum of the homopolymers mixture, allows us to rule out the possibility that the microgel structure consists of two distinct blocks, one formed exclusively by PNIPAM and the other exclusively by PNIPMAM. Nonetheless, among these six peaks, the contributions corresponding to the homopolymer signals can still be identified at around 175.2 ppm and 178.0 ppm, indicating the presence of homopolymeric NIPAM and NIPMAM sequences, respectively.
These have been labelled in Figure~\ref{fig:triad}d as PPP and MMM, in agreement with the earlier discussion. In between these two signals, the other possible combinations of each repeating unit linked to a different neighboring unit can be observed: PPM at approximately 175.8 ppm, MPM at approximately 176.1 ppm, PMP at approximately 176.9 ppm, and MMP at approximately 177.4 ppm.
In the assignment, possible additional peaks due to deuterated units were neglected, as they were assumed to be almost superimposed on those of the hydrogenated ones. This assumption is supported by the absence of hydrogens or deuterons directly bonded to the analyzed carbon, and by the lack of other clearly distinguishable contributions besides the six ones marked in Figure~\ref{fig:triad}d.
The deconvolution of these signals allowed us to determine the contribution of each sequence to the overall peak, with their respective percentage areas reported in Figure~\ref{fig:triad}e. It is evident that NIPAM has a higher tendency to form block-like sequences as compared to NIPMAM, since the PPP contribution is consistently larger than the MMM contribution. Conversely, NIPMAM shows a preference for bonding with NIPAM repeating units, leading predominantly to the formation of MMP or PMP sequences. This distribution is observed in all three copolymers, thus demonstrating a consistent pattern of bond formation within the microgel.

\section{Conclusions}
In this work, we have put forward a comprehensive investigation, integrating small-angle neutron scattering, DLS and NMR measurements with multi-scale numerical simulations, to unveil the internal architecture of copolymer microgels, specifically of P(NIPAM-\textit{co}-NIPMAM) type. To this aim, we synthesized copolymer microgels at equimolar composition with different isotopic composition (hydrogen/deuterium), allowing us to distinguish PNIPAM and PNIPMAM contributions via SANS experiments. The measured form factors were then directly compared to distinct copolymer topologies in monomer-resolved simulation, which allowed us to identify the
\emph{block} topology as the most accurate representation of the experimental scattering profiles.
We stress that, while also a \emph{random} topology would be satisfactory for P(H-NIPAM-\textit{co}-H-NIPMAM), it is the use of selective scattering that has enabled us to distinguish the subtle difference between the two arrangements.  These results support the formation of local domain structures within the microgel network, suggesting that repeating units of the same type (either NIPAM or NIPMAM) preferentially cluster together rather than being randomly and homogeneously distributed.
This behavior can be rationalized by considering the different polymerization kinetics of the two monomers: specifically, NIPAM is known to exhibit a higher propagation rate than NIPMAM~\cite{duracher1999preparation}, favoring the formation of contiguous sequences of the same polymer during synthesis.
However, this is the first report of this kind in the literature of copolymer microgels synthesized via a one-step precipitation polymerization process.
For this reason, we challenged them very seriously by (i) performing atomistic molecular dynamics simulations of isolated copolymer chains and (ii) carrying out careful $^{13}$C-NMR measurements of the microgels in comparison to the respective homopolymers. These two additional investigations reinforce the hypothesis of block formation. Indeed, simulations clearly demonstrate that a \emph{block} copolymer configuration more accurately captures the coil-to-globule transition as compared to a \emph{random} model. Finally, the close inspection of the carbonyl peak of the  $^{13}$C-NMR spectra shows clear evidence of NIPAM block formation, while NIPMAM is organized in a more random structural environment. The asymmetry of block formation  is present in all microgel samples, including deuterated ones, providing robustness to the  results. This observation is well compatible with the larger reactivity of NIPAM which is more prone to form the blocks within the polymerization reaction. Indeed, the most realistic microstructure of polymer chains exhibits a gradient composition, beginning with block-like PNIPAM sequences and gradually transitioning to a more random P(NIPAM-r-NIPMAM) distribution. This evolution results from the increasing concentration of the less reactive monomers in the reaction medium. Although NIPMAM monomer is inherently less reactive, its growing excess as the reaction progresses increases its probability of incorporation into the polymer chain.
Interestingly, the atomistic approach additionally provides a clear molecular pictute of the consequences of this domain formation, highlighting a pronounced dependence of polymer-water hydrogen bonding on the local monomeric sequence. Specifically, the ability of individual monomers to form hydrogen bonds with surrounding water molecules is significantly influenced by their neighboring units, thereby favoring the formation of contiguous regions composed of the same monomer type within the copolymer network.

While our analysis provides evidence for the presence of compositional heterogeneities, the precise size and spatial extent of these local domains remain unresolved. The determination of the domain size is beyond the resolution of SANS form factor analysis and NMR capability, thus representing a challenging direction for future experimental efforts, potentially requiring complementary techniques with higher spatial resolution. To verify what happens at the synthesis level, it would also be useful to investigate the kinetics of the two monomers consumption, as well as the exact microstructure. Surely, the present results encourage the investigation of other copolymer microgels by similar analysis of the internal structure to probe the generality of such a feature when mixing co-monomers with different polymerization reactions. It would thus be important to re-assess copolymer microgels made of NIPAM/NIPMAM and VCL~\cite{balaceanu2013copolymer,wu2014behavior} with the present understanding to verify whether this phenomenon is of a more general nature and may be found in the majority of copolymer microgels.

In summary, the present study has the potential to radically shift the common understanding and interpretation of copolymer microgels from a simple random mixing of the two polymers to a more structured type of material, that could be promising for enhancing the control of material responsivity and functional behavior.

\section{Models and methods}

\subsection{Synthesis of copolymer microgels}
Equimolar copolymer microgels with 5\% crosslinker and 1\% initiatior were synthesized via precipitation polymerization \cite{fernandez2011}, using N-iso\-propyl\-acryl\-amide (NIPAM, $\text{MW}=113.16$ Da, Sigma, 97\% purity) and N-iso\-propyl\-methacryl\-amide (NIPMAM, $\text{MW}=127.18$ Da, Sigma, 97\% purity) monomers, N,N'-methylene\-bis\-acryl\-amide (BIS, $\text{MW}=154.17$ Da, Sigma, 99\% purity) crosslinker, Sodium dodecyl sulfate (SDS, $\text{MW}=288.38$ Da, Sigma, 99\% purity) surfactant and potassium persulfate (KPS, $\text{MW}=270.32$ Da, Sigma, 99\% purity) initiator.
Briefly, all reagents except the initiator are dissolved in 26.5 mL of deionized water to yield final concentrations of 160 mM monomer, 8 mM crosslinker, and 4 mM surfactant.
The same molar concentrations are used to synthesize microgels with deuterated monomers, for which  NIPAM-d$_{10}$ and NIPMAM-d$_{12}$ monomers were synthesized at the J\"ulich Centre for Neutron Scattering according to the procedures reported in the SI.
In these cases, the monomer compositions of the microgels are adjusted to have the same scattering length density in SANS experiments, as follows: 16 mol\% H-NIPAM and 34 mol\% D-NIPAM (NIPAM-d$_{10}$) for D-H sample, and 22 mol\% H-NIPMAM and 28 mol\% D-NIPMAM (NIPMAM-d$_{12}$) for H-D one.
The solution is loaded into a 50 mL two-necked reactor equipped with a condenser and a magnetic stirrer, and  thermostated using an oil bath. The solution is then bubbled under a nitrogen stream for 1 hour at room temperature. Separately, the initiator is dissolved in degassed water at the concentration of 36.9 mM. Subsequently, the temperature in the reactor is raised to 70°C and the polymerization reaction is initiated by adding 1.2 mL of the initiator solution at the rate of 1 mL/min. The final concentration of the initiator in the reaction solution is 1.6 mM. The reaction is carried out at constant temperature for 5 hours. The resulting microgels are purified by dialysis against ultrapure water using a cellulose membrane (6–8 kDa MWCO, Sigma) for two weeks, with the water replaced twice daily. The microgel solutions are then freeze-dried and stored in the dark at 4°C.

\subsection{SANS}

SANS measurements were performed at QUOKKA (ANSTO, Sydney, Australia)\cite{Wood2018} and at SANS-I (PSI,
Villigen, Swiss)\cite{kohlbrecher2000}, using the following configurations. For QUOKKA: {\it(i)} 1.35 m Sample-to-Detector Distance (SDD) and incident wavelength $\lambda$ = 5.5 \AA, {\it(ii)} 12 m SDD and $\lambda$ = 5.5 \AA, and {\it(iii)} 20 m SDD and $\lambda$ = 8.1 \AA.
The combination of the three configurations gives a wave vector range 0.0007 $\text{\AA}^{-1} < Q <$ 0.66$\text{ \AA}^{-1}$. For SANS-I: we used a constant $\lambda$ = 8 \AA~for two SDD of: 4.5 m and 18 m. The combination of the two configurations gives a wave vector range 0.0024 $\text{\AA}^{-1} < Q <$ 0.15$\text{ \AA}^{-1}$.
For data measured at Quokka a macro in Igor Pro software (Wavemetrics, Lake Oswego, Oregon, USA), originally written by Kline\cite{Kline2006}, was used for data reduction, while BerSANS\cite{keiderling2002} was used at SANS-I.
Dilute sample suspensions (0.1 wt\% for D-H; 1\% for H-H and H-D) were measured at different temperatures between 20°C and 50°C in quartz cells with a path length of 2~mm (Hellma GmbH \& Co., Mullheim, Germany).
After each temperature variation, samples are kept thermalizing for at least 5 minutes.
All scattering data were normalized for the sample transmission and background corrected using a quartz cell filled with D$_2$O. All data were analyzed within the SASView package using standard and user-written functions.\\
The particle form factor in SANS experiments was described using  the fuzzy sphere model from Stieger et al. \cite{Stieger2004}, which is able to reproduce the data under all studied conditions. No structure factor contributions were detected at the concentration here investigated. The form factor model reads as,
\begin{equation}\label{eq:Pfuzzy}
P(Q)= A_1\left[ \frac{3[\sin(QR) - QR\cos(QR)]}{(QR)^3} \cdot \exp{\left( - \frac{(\sigma Q)^2}{2} \right)} \right]^2 + A_{2}\frac{1}{1 + (Q\zeta)^2},
\end{equation}
in which the first term represents the fuzzy sphere contribution, being R the radius of the sphere and $\sigma$ the fuzziness parameter, while the second term describes the polymer network scattering in the region at high Q, with $\zeta$ the correlation length of the polymer mesh. For the fits, the model function was convoluted with the following experimental smearing function:
\begin{equation} \label{eq:smear}
    I(q)= \int P(q-q') \left( \frac{1}{2\pi \sigma_{q'}^{2}} \right)^{1/2} \exp{ \left[ - \frac{q'^{2}}{2 \sigma_{q'}^{2}}  \right]}   \text{d}q'
\end{equation}
where $\sigma_{q}$ is the standard deviation of the $q$~resolution, which contains the detector resolution and the beam wavelength spread contributions\cite{pedersen1993}. In addition, sample polydispersity was included considering a Schulz distribution of the radius. Obtained values of the sample polydispersity are included in Tab.S2 of the SI. To facilitate the direct comparison with simulations, experimental data in Fig.~\ref{fig:models-ff} and Fig.~\ref{fig:ffnew} refer to the fit results via Eq.~\ref{eq:Pfuzzy}, where the resolution smearing and polydispersity effects have been subtracted. An exemplary form factor with fit including resolution smearing and polydispersity and after removal of resolution smearing and polydispersity is reported in Figure S17.

\subsection{Dynamic light scattering}
Hydrodynamic radius ($R_H$) distributions are measured by DLS using a NanoZetaSizer apparatus (Malvern Instruments LTD) equipped with a \ch{He-Ne} laser (5 mW power, 633 nm wavelength). The scattered light is collected in quasi-backscattering geometry (at an angle of 173$^{\circ}$).
The distribution of diffusion coefficients $D$ of the particles is derived by extrapolating the decay times from the acquired intensity autocorrelation functions. Diffusion coefficients are then converted to intensity-weighted distributions of $R_H$ using the Stokes-Einstein relationship $R_H = k_BT/6\pi\eta D$, where $k_BT$ is the thermal energy and $\eta$ the water viscosity.
Temperature trends are acquired between 20°C and 50°C with 1°C steps, by keeping the samples thermalizing for 5 minutes after each temperature variation before performing the measurement.
The values of $R_H$ reported in the work and the associated errors are the average and standard deviation, respectively, of a distribution obtained by at least 30 measurements.

\subsection{NMR}
$^1$H-NMR and $^{13}$C-NMR spectra were recorded by NMR Ascend 500 MHz Bruker instrument equipped with iProbe BBFO 5mm BBF/1H probe and autosampler, at room temperature. For measurements we used a sample concentration of 20-30 mg/mL in deuterium oxide. For the $^{13}$C-NMR analysis, the impulse angle was of 30°, the acquisition time of 0.7 s, the delay of 0.1 s, number of scans of 200000 and temperature of acquisition of 25°C.\\
Following Ref.~\citenum{Friesen1999}, the fraction of PNIPAM $f_\text{NIPAM}$ in the H-H copolymer microgel was calculated in the $^1$H-NMR spectrum from the the area under the peaks in the range $0.5 - 2.5$ ppm (see Figure \ref{fig:Exp_FF} for assignment, where the peaks are labelled as b, c, d and e, respectively) using the equation below:
\begin{equation}\label{eq:NMRratio}
  f_\text{NIPAM} (\%) = \frac{A - A_\text{PNIPMAM}}{A_\text{PNIPAM} - A_\text{PNIPMAM}} \cdot 100
\end{equation}
where $A$, $A_\text{PNIPMAM}$ and $A_\text{PNIPAM}$ denote the area under the spectrum in the region $0.5 - 2.5$ ppm for the H-H copolymer microgel, homopolymer PNIPMAM microgels, and homopolymer PNIPAM microgels, respectively.

In the $^{13}$C-NMR spectra, the deconvolution of the peaks was performed using MestReNova software, as follows. Initially, the peak chemical shifts were fixed at the exact values. Then, the fitting procedure, using a Lorentzian peak shape, was performed allowing a maximum shift of ±0.2 ppm. The fit was considered acceptable when the sum of the contributions of the NIPAM peaks (PPP, PPM, and MPM) matched the sum of the NIPMAM peaks (MMM, MMP, and PMP), in agreement with the molar ratio used in the microgel synthesis. The relative percentage contribution of each sequence to the overall peak was then determined.

\subsection{\emph{In silico} investigation of copolymer microgels}
We employ a monomer-resolved model of copolymer microgels composed by $N\sim42000$ beads, that is based on a recently developed \emph{in silico} synthesis protocol~\cite{gnan2017silico}. This procedure leads to the formation of fully bonded, disordered polymer networks that were shown to well reproduce the experimental swelling behavior of pure PNIPAM microgels~\cite{ninarello2019modeling,hazra2023structure}. To extend it to copolymer microgels, NIPAM and NIPMAM repeating units are both described by bivalent patchy particles, while cross-linkers have four attractive patches. The self-assembly of these particles is carried out in a confining sphere under the influence of an external force acting on the crosslinkers only, which drives the formation of a fuzzy sphere structure~\cite{ninarello2019modeling}.
In order to reproduce the experimental synthesis conditions, the crosslinker concentration is set to $c$ = 5.0\% and the copolymer ratio is 50\%.

The microgels polymer network is described by beads interacting with the Kremer-Grest potential~\cite{grest1986molecular}. Therefore, beads experience a steric repulsion, modelled by the Weeks-Chandler-Anderson (WCA) potential:
\begin{equation}\label{WCA}
  V_\text{WCA}(r)=\begin{cases} 4\varepsilon \left[\left(\frac{\sigma}{r}\right)^{12}-\left(\frac{\sigma}{r}\right)^6\right]+\varepsilon & \qquad \textrm{if}\quad r\leq2^{\frac{1}{6}}\sigma \\
  0 & \qquad \textrm{if}\quad r>2^{\frac{1}{6}}\sigma \\
  \end{cases}
\end{equation}
where $\epsilon$ and $\sigma$ are the energy and length units, respectively. In addition, bonded beads interact via the finitely extensible nonlinear elastic potential (FENE):
\begin{equation}\label{FENE}
  V_\text{FENE}(r)=-\varepsilon k_F R_0^2 \,\log \left[ 1- \left( \frac{r}{R_0\sigma} \right)^2 \right] \qquad r<R_0\sigma
\end{equation}
where $R_0\sigma$ is the maximum bond distance and $k_F$ is a stiffness parameter that determines the bond rigidity. The covalent nature of the network is mimicked  by making bonds that cannot break during the simulation.

The solvent contribution is implicitly included through an effective potential, which mimics the change in the polymer-solvent affinity by raising temperature:

\begin{equation}\label{alpha}
  V_{\alpha}(r)=\begin{cases} -\varepsilon\alpha & \qquad \textrm{if}\quad r\leq2^{\frac{1}{6}}\sigma \\
  \frac{1}{2}\alpha\varepsilon \left\{ \cos \left[ \gamma \left( \frac{r}{\sigma} \right)^2 + \beta \right] -1 \right\} & \qquad \textrm{if}\quad 2^{\frac{1}{6}}\sigma < r \leq R_0\sigma \\
  0 & \qquad \textrm{if}\quad r > R_0\sigma
  \end{cases}
\end{equation}

where $\alpha$ is the solvophobicity parameter corresponding to the effective temperature. The swelling curve of the microgels is obtaining by changing the value of $\alpha$ from good ($\alpha=0$) to bad ($\alpha=1.5$) solvent conditions. Here $\gamma = \pi \left (\frac{9}{4} - 2^\frac{1}{3} \right)^{-1}$ and $\beta=2\pi-\frac{9}{4}\gamma$ are constants defining the functional shape of the potential\cite{soddemann2001generic}.

The four microgel models described in the main text are considered, assigning different chemical nature of the NIPAM and NIPMAM repeating units. First of all, the bead size is set to $\sigma$ = 1 for the former and $\sigma$ = 1.04 for the latter to account for their steric volume difference. In addition, the different volume phase transition temperature of 305 K and 316 K is realized for PNIPAM and PNIPMAM microgels, respectively, by assigning a different value of $\alpha$ to each monomer, based on the relation between $\alpha$ and temperature determined for PNIPAM microgels in Ref.~\citenum{gnan2017silico}, as illustrated in Figure~SX. For mixed interactions occurring between PNIPAM and PNIPMAM beads the average value of $\alpha$ is used. The value of $\alpha$ reported in the manuscript always refers to the PNIPAM $\alpha$-value for convenience.

The equations of motion are integrated through a Nos\'{e}–Hoover thermostat in the constant NVT ensemble with an integration time-step $\Delta t = 0.002\tau$, where $\tau=\sqrt{\frac{m\sigma^2}{\varepsilon}}$ is the reduced time unit. Simulations are performed with the LAMMPS package~\cite{plimpton1995fast} in a cubic box with side L = 200 $\sigma$ at a fixed temperature $\frac{k_BT}{\epsilon}=1.0$. The equilibration of each system is carried out for 1$\cdot$10$^6\tau$, followed by a production run of 1$\cdot$10$^7\tau$.

To directly compare numerical results to the experimental measurements, we computed the microgel form factor $P(q)$ and hydrodynamic radius $R_H$. $P(q)$ is given by
\begin{equation}\label{FF}
  P(q)=\frac{1}{N}\sum_{ij}\langle exp (-i\vec{q}\cdot\vec{r}_{ij}) \rangle
\end{equation}
where $q$ is the wavevector, $r_{ij}$ is the distance between the monomers $i$ and $j$ and the calculation is averaged over independent configurations. In the case of partially deuterated samples, the form factor has been calculated only for the beads belonging to the non-deuterated polymer, due to the high difference in scattering between hydrogenated or deuterated monomers. The hydrodynamic radius $R_H$ is calculated as:\cite{hubbard1993,del2021two}
\begin{equation}
    R_H 
 =2\left[ \int_0^\infty \frac{1}{\sqrt{(a^2+\theta)(b^2+\theta)(c^2+\theta)}} d\theta \right]^{-1}
\end{equation}
where the microgel is approximated as an effective ellipsoid with
principal semiaxes $a$, $b$ and $c$.

\subsection{Atomistic molecular dynamics simulations of copolymer chains}
All-atom molecular dynamics simulations are performed on a model single chain composed by 90 repeating units in aqueous solution at infinite dilution. We consider pure PNIPAM, pure PNIPMAM and two different copolymer models with a composition ratio of 50\%: (i) a random model where PNIPAM and PNIPAM repeating units are randomly distributed; and (ii) a block model in which the chain is composed by two separate blocks of 45 repeating units. For all systems, the stereochemistry of the polymer chains is set as atactic. PNIPAM and PNIPMAM are described with the force field OPLS-AA~\cite{jorgensen1996development}, applying the modifications of Siu et al.~\cite{siu2012optimization}. TIP4P-Ice model is used for water~\cite{abascal2005potential}. This computing setup has been effective in reproducing the pressure/temperature phase behaviour of PNIPAM and PNIPMAM aqueous solution~\cite{tavagnacco2021molecular,tavagnacco2022modeling,tavagnacco2025understanding}.

NPT MD simulations are performed at 0.1 MPa at different temperatures from 280 to 320 K. Trajectories of 1 $\mu$s are acquired at each temperature, in two independent replicas, using the GROMACS software package (versions 2020.3 and 2022.3)~\cite{abraham2015gromacs,markidis2015solving}. The last 500 ns interval is used as the production run, sampling 1 frame every 100 ps. The leapfrog integration algorithm~\cite{hockney1970potenitial} was used with a time step of 2 fs. Periodic boundary conditions and minimum image convention were applied. The length of bonds involving H atoms was constrained using the LINCS procedure~\cite{hess1997lincs} The velocity rescaling thermostat coupling algorithm with a time constant of 0.1 ps was used to control temperature.~\cite{bussi2007canonical} Pressure was maintained using the Parrinello–Rahman approach, with a time constant of 2 ps.~\cite{parrinello1981polymorphic,nose1983constant} The cutoff of nonbonded interactions was set to 1 nm, and electrostatic interactions were calculated using the smooth particle-mesh Ewald method.~\cite{essmann1995smooth}

\begin{acknowledgement}
We thank F. Camerin for early discussions on this project. EB, EC, ML and EZ acknowledge financial support by Progetto Co-MGELS funded by the European Union - NextGeneration EU under the National Recovery and Resilience Plan (PNRR) Mission 4 “Istruzione e Ricerca” - Component C2 - Investment 1.1 - "Fondo PRIN", Project code PRIN2022PAYLXW Sector PE11, CUP B53D23008890006. LT and EZ also acknowledge financial support from ICSC-Centro Nazionale di Ricerca in High-Performance Computing, Big Data and Quantum Computing-(Grant No. CN00000013, CUP J93C22000540006, PNRR Investimento M4.C2.1.4). JV acknowledges funding from Ministero dell'Università e della Ricerca (D.D. 247 published on 19.08.2022, grant Nr. MSCA\_0000004), funded by European Union - NextGenerationEU - under the National Recovery and Resilience Plan (PNRR), Mission 4, Component 2, Investment line 1.2.
The CINECA award, under the ISCRA initiative, is acknowledged for the availability of high-performance computing resources and support.
\end{acknowledgement}

{}

\clearpage
\newpage
\onecolumn
\setcounter{figure}{0}
\setcounter{section}{0}
\setcounter{page}{1}
\renewcommand{\theequation}{S\arabic{equation}}
\renewcommand{\thefigure}{S\arabic{figure}}
\renewcommand{\thetable}{S\arabic{table}}
\renewcommand{\thepage}{S\arabic{page}}
\renewcommand{\theequation}{S\arabic{equation}}
\renewcommand{\bibnumfmt}[1]{[S#1]}
\renewcommand{\citenumfont}[1]{S#1}

\section{Supporting Information}

\section{Additional Methods}
\subsection{Synthesis of deuterated monomers}
\subsubsection{NIPAM-d$_{10}$}
All chemicals were used as received. CHCl$_3$, n-hexane and toluene (all absolute or HPLC grade) as well as NaCl were purchased from VWR. D$_2$SO$_4$ (98\%, 99.5\% D) and isopropanol-d$_8$ (99,5\% D) were purchased from Armar Isotopes. MgSO$_4$ (anhydrous) was purchased from Th.Geyer. Acrylonitrile-d$_3$ (98.7\% D) and D$_2$O (99.90\% D) were purchased from Eurisotop.\\
D$_2$SO$_4$ (conc., 50 mL) was cooled to 0 °C in a flask covered with aluminum foil. Then acrylonitrile-d$_3$ (10.0 g, 11.6 mL, 178 mmol, 1.0 equiv.) was added and the solution was stirred for 5 minutes before slowly adding 2-propanol-d$_8$ (24.3 g, 27.3 mL, 356 mmol, 2.0 equiv.) via a dropping funnel over 1 hour. The ice bath was removed and the mixture stirred at room temperature for 3 h before pouring it onto 400 mL of ice. The aqueous mixture was slowly neutralized with solid Na$_2$CO$_3$, saturated with NaCl and extracted with CHCl$_3$ until TLC analysis (SiO$_2$, EtOAc) showed, that all of the product had moved into the organic phase. The organic phase was dried over MgSO$_4$, stabilized with a spatula tip of BHT and dried in vacuo. The crude product was then purified via sublimation (~55 °C water bath, 2$\cdotp$10$^{-3}$ mbar vacuum) to yield 16.2 g (131 mmol, 74\%) of NIPAM-d$_{10}$. An additional purification via recrystallization from a mixture of n-hexane and toluene yielded 13.5 g (109 mmol, 61\%) of clean product as colorless crystals.
\subsubsection{Characterization}
The obtained materials were characterized by NMR. NMR spectra were collected on a Varian INOVA 400 MHz spectrometer. All samples were measured at 295 K. Samples were either diluted in CDCl$_3$ (for 1H-NMR) or CHCl$_3$ (for 2H-NMR). 1,2,4,5-Tetrabromobenzene was used as an internal standard to determine the deuteration degree and purity of the products to be > 97\%.

\subsubsection{NIPMAM-d$_{12}$}
All chemicals were used as received. Isopropanol-d$_8$ (99.5\% D), KBr ($\geq$ 99.0\%), potassium phthalimide ($\geq$ 99.0\%), DMF (anhydrous, 99.8\%), hydrazine monohydrate (98\%), MeOH ($\geq$ 99.8\%), KOH ($\geq$ 85\%), Dichloromethane ($\geq$ 99.9\%), 1,2,4,5-tetrabromobenzene (97\%) and thionyl chloride ($\geq$ 99\%) were purchased from Sigma-Aldrich.
\ch{H2SO4} (95-97\%), HCl ($\geq$ 37\%) and MgSO$-4$ were purchased from Merck. CHCl$_3$, EtOH, Et$_2$O, n-hexane and toluene (all absolute or HPLC grade) were purchased from VWR. Methyl methacrylate-d8 (99.5\% D) was purchased from Armar Isotopes. The synthesis was composed of the following steps.

\subsubsection{Synthesis of Isopropyl Bromide-d$_7$}
Isopropanol-d$_8$ (100 g, 1.467 mol, M=68.14 g/mol) and KBr (181 g, 1.521 mol, M=119 g/mol) were dissolved in 135 ml H$_2$O in a 500 ml round bottom flask equipped with a strong magnetic stir bar, a distillation head and a reflux condenser. 198 ml conc. sulfuric acid were added to the reaction mixture at 0 °C via the condenser. After the addition was finished the reaction was heated to 70 °C. Once the temperature in the gas phase had reached 50 °C the product was collected in a 250 ml cooled Schlenk bomb. The bath temperature for the distillation was slowly raised to 90 °C and the reaction was continued until no more product was collected. 190.64 g of a colorless liquid were collected (93.6\% crude yield) 2H-NMR showed ~5\% side product formation. The crude product was used in the next step without further purification.
\subsubsection{Synthesis of N-Isopropyl-d$_7$ Phthalimide}
Potassium phthalimide (185.2 g, 1.374 mol 1 eq., M=185.22 g/mol) and crude 2-bromopropane-d$_7$ (equivalent to 178.5 g, 1.374 mol, M=129.9 g/mol based on 2-bromopropane content of the crude product) were dissolved in 700 ml DMF in a 1 L round bottom flask equipped with a large magnetic stir bar and a reflux condenser. The mixture was stirred for 1 h at 70 °C, 1 h at 120 °C and 2 h at 150 °C. The resulting solution was cooled in a fridge overnight, filtered and the filtrate was washed with 200 ml cold DMF. The organic solutions were combined and the solvent was removed under reduced pressure to yield the crude product as a yellow solid. Further drying under high vacuum conditions yielded 258.0 g crude product. The crude product was dissolved in 300 ml CHCl$_3$ and filtered for further purification. After removal of the solvent and drying 234.9 g product remained (87.2\% chemical yield, 81.7\% total yield over 2 steps).
\subsubsection{Synthesis of Isopropylamine-d$_7$} N-Isopropyl-d$_7$ phthalimide (20.0 g, 102 mmol, 1.0 equiv) was suspended in 30 mL of methanol. Hydrazine monohydrate (5.2 mL, 107 mmol, 1.05 equiv) was added and the mixture was heated to 80 °C for 1 h. The volatile components were removed in vacuo and the solid residue was cooled to 0 °C and treated with 30 mL of a 7.4 M aqueous KOH solution. The flask was fitted with a Vigreux column and a distillation head and the product was distilled off while slowly heating the oil bath from 75 °C to 110 °C, yielding 5.79 g (87.6 mmol, 86\%) of isopropylamine-d$_7$ as a colorless liquid.
\subsubsection{Synthesis of Methacrylic Acid-d$_5$}
Methyl methacrylate-d$_8$ (13.0 g, 120 mmol, 1.0 equiv) was dissolved in 120 mL of a mixture of DCM and EtOH (9:1). Then 24 mL of a 3 N methanolic KOH solution was added and the reaction was stirred at room temperature for 1 h. The solvent was evaporated in vacuo and the residue was dissolved in water and extracted with DCM and Et2O (1x100 mL each). The aqueous phase was treated with 1 M HCl until the pH reached 2 and then extracted with Et2O (2x) and DCM. The combined organic phases were stabilized by adding hydroquinone (100 mg), dried over MgSO4 and the solvent was evaporated in vacuo (min. 100 mbar, 40 °C) to yield 5.73 g (60 mmol, 50\%, containing 0.28 g of EtOH) of methacrylic acid-d$_5$ as a colorless liquid that was used in the next step without further purification.
\subsubsection{Synthesis of NIPMAM-d$_{12}$}
Crude methacrylic acid (KS058, 60 mmol, 1.0 equiv) was dissolved in 12 mL of DCM alongside 0.2 mL of DMF as a catalyst. Thionyl chloride (4.6 mL, 63 mmol, 1.05 equiv) was added dropwise and the mixture was stirred under argon for 4 h. In a second flask isopropylamine-d7 (4.1 g, 62 mmol, 1.03 equiv) and triethylamine (17.6 mL, 126 mmol, 2.1 equiv) were dissolved in 120 mL of DCM cooled to 0 °C. Then the crude mixture of acid chloride and DCM was added dropwise. The resulting mixture was stirred over night while slowly warming up to room temperature. Then 100 mL of water were added and the mixture was extracted with DCM (2 x 100 mL). The combined organic phases were washed with 1 M aqueous HCl (2 x 100 mL) and brine (100 mL), dried over \ch{MgSO4} and the solvent was evaporated in vacuo. The crude product was recrystallized from n-hexane and toluene to yield 4.5 g (32 mmol, 53\%) of NIPMAM-d$_{12}$ as colorless crystals. The deuteration degree was determined by 1H-NMR using 1,2,4,5-tetrabromobenzene as a standard to be >98\%.
\subsubsection{NMR characterization}
NMR spectra were collected on a Bruker Avance III 600 MHz spectrometer, equipped with a Prodigy cryoprobe with a 5 mm PFG AutoX DB probe.
All samples were measured at 295 K. Samples were either diluted in \ch{CDCl3} (for 1H-NMR) or \ch{CHCl3} (for 2H-NMR). We used 1,2,4,5-Tetrabromobenzene as internal standard to determine the deuteration degree and purity of the products to be > 98\%.
\newpage

\section{Additional Results}
\subsection{Swelling curves of P(NIPAM-\textit{co}-NIPMAM) microgels from dynamic light scattering and monomer-resolved simulations}
\begin{figure*}
\centering
\includegraphics[width=0.8\linewidth]{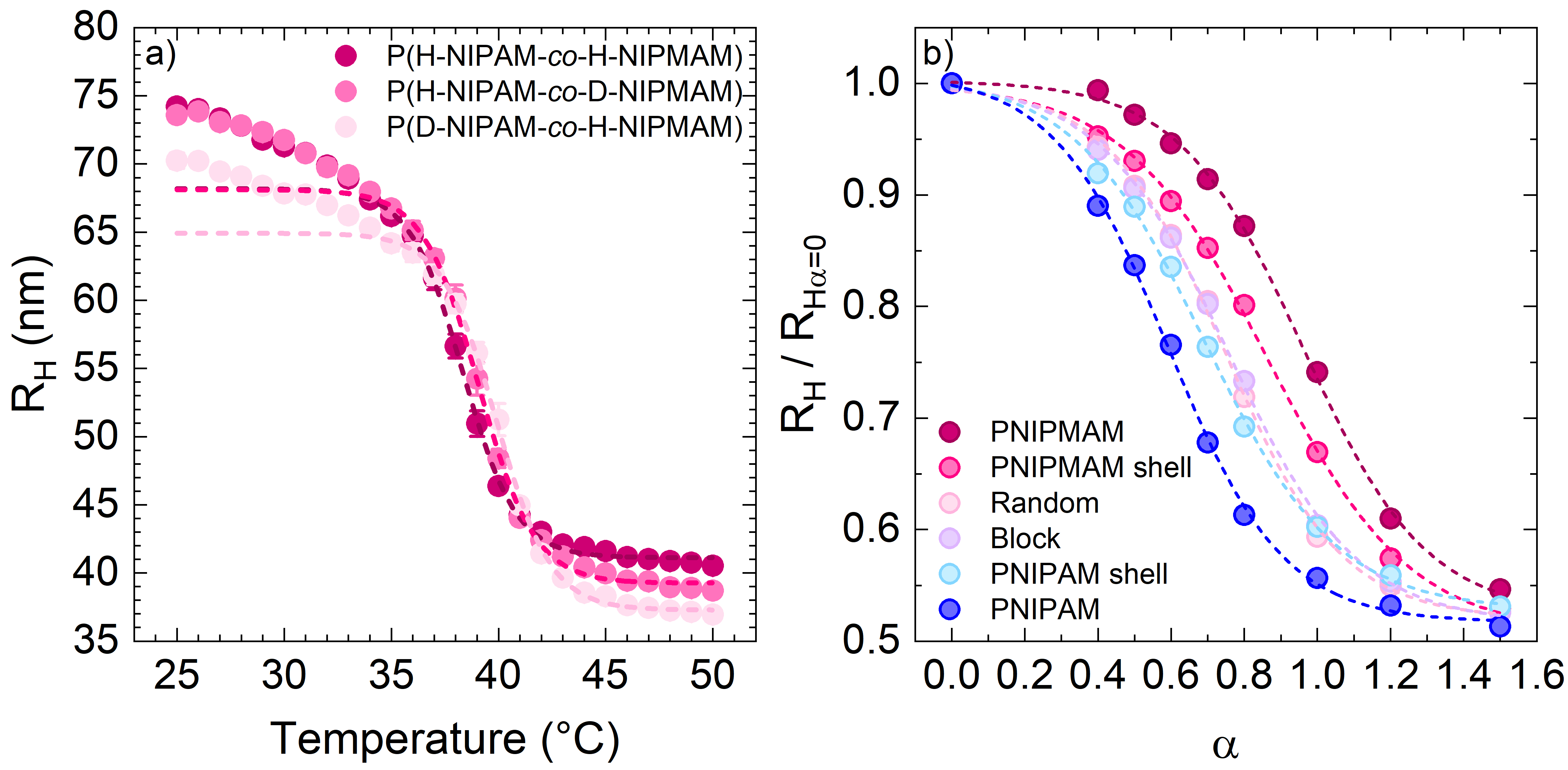}
\caption{a) Hydrodynamic radius ($R_H$) from DLS measurements as a function of temperature for H-H (red), H-D (magenta) and D-H (pink). b) Hydrodynamic radius normalized to the corresponding value at the solvophobic parameter $\alpha=0$ as a function of $\alpha$ for the different microgel topologies: PNIPMAM shell (magenta circles), block (purple circles), random (pink circles), PNIPAM shell (light blue circles), and pure PNIPAM (blue circles) and PNIPMAM (red circles). The dashed lines are sigmoidal fits; for experimental data we set the fit range in proximity of the VPT.}
\label{fig:swelling}
\end{figure*}

The swelling curves of the P(NIPAM-\textit{co}-NIPMAM) microgels from experiments in H$_2$O and monomer-resolved simulations are shown in Figure~\ref{fig:swelling} a and b, respectively. The curves are fitted to a sigmoidal function, obtaining $T_\text{VPT} = 38.4 \pm 0.1\degree$C for H-H, $T_\text{VPT} = 39.1 \pm 0.1\degree$C for H-D and $T_\text{VPT} = 39.9 \pm 0.1\degree$C for D-H. The fitting parameters of the simulations are reported in Table~\ref{tbl:TVPT}.

\begin{table}
  \caption{$T_\text{VPT}$ from the sigmoidal fit of the numerical swelling curves.}
  \label{tbl:TVPT}
  \setlength{\tabcolsep}{24pt}
  \centering
  \begin{tabular}{lcc}
   \hline
    Microgel topology	    & $\alpha_\text{VPT}$   &  $T_\text{VPT}$ / K\\
   \hline
   PNIPMAM		   & 0.96 ($\pm$0.01) & 317.0 \\
   PNIPMAM shell   & 0.86 ($\pm$0.01) & 313.3 \\
   Random          & 0.75 ($\pm$0.01) & 308.9 \\
   Block           & 0.76 ($\pm$0.01) & 309.2 \\
   PNIPAM shell    & 0.69 ($\pm$0.01) & 306.8 \\
   PNIPAM          & 0.59 ($\pm$0.01) & 303.0 \\
   \hline
   \end{tabular}
\end{table}

\clearpage
\newpage
\subsection{$^1$H-NMR of microgels}
\begin{figure*}
\centering
\includegraphics[width=0.8\linewidth]{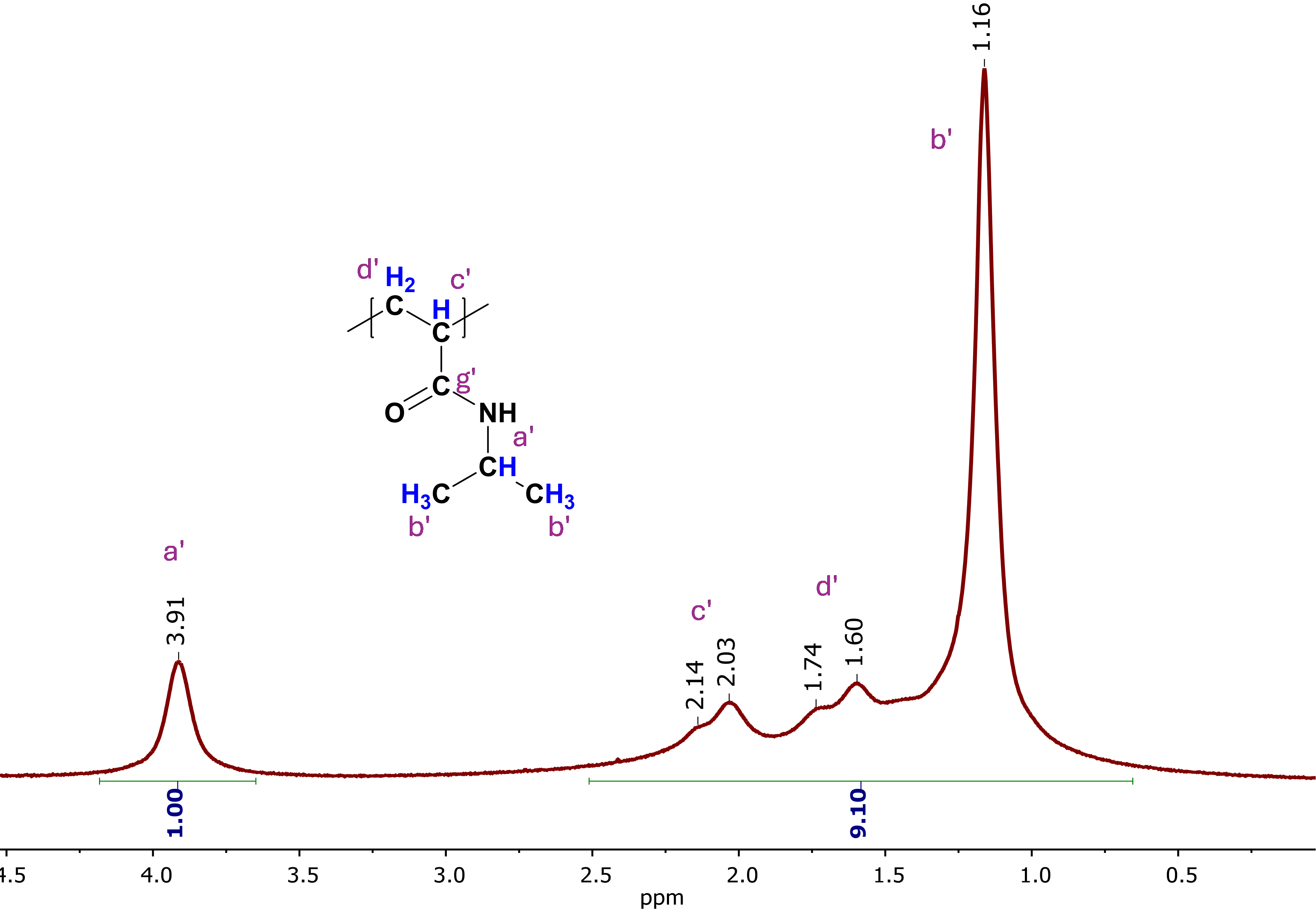}
\caption{$^1$H-NMR spectrum at 25°C of PNIPAM microgel dispersion in D$_2$O.}
\label{fig:SI_NMR_1}
\end{figure*}

\begin{figure*}
\centering
\includegraphics[width=0.8\linewidth]{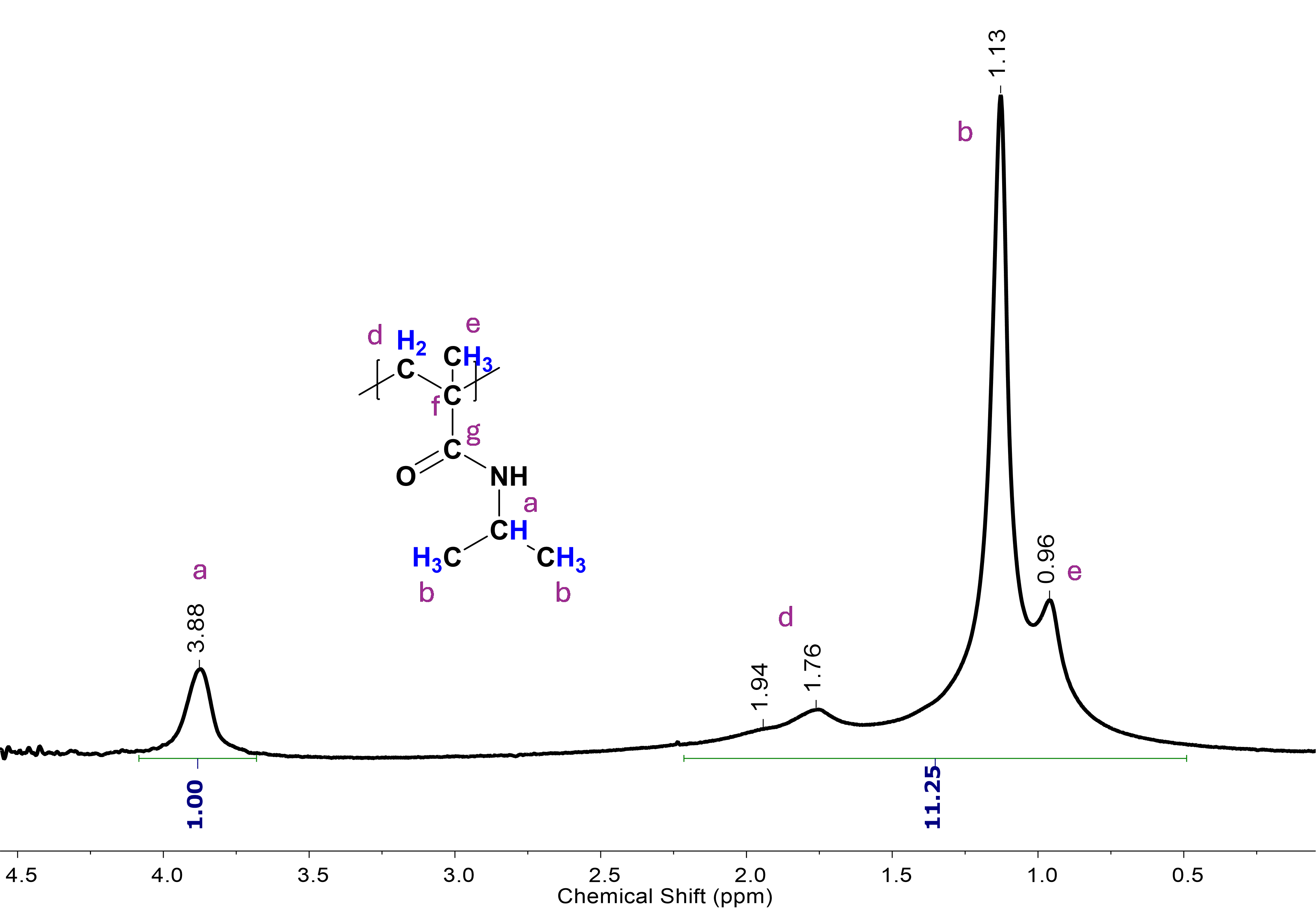}
\caption{$^1$H-NMR spectrum at 25°C of PNIPMAM microgel dispersion in D$_2$O.}
\label{fig:SI_NMR_2}
\end{figure*}

\begin{figure*}
\centering
\includegraphics[width=0.8\linewidth]{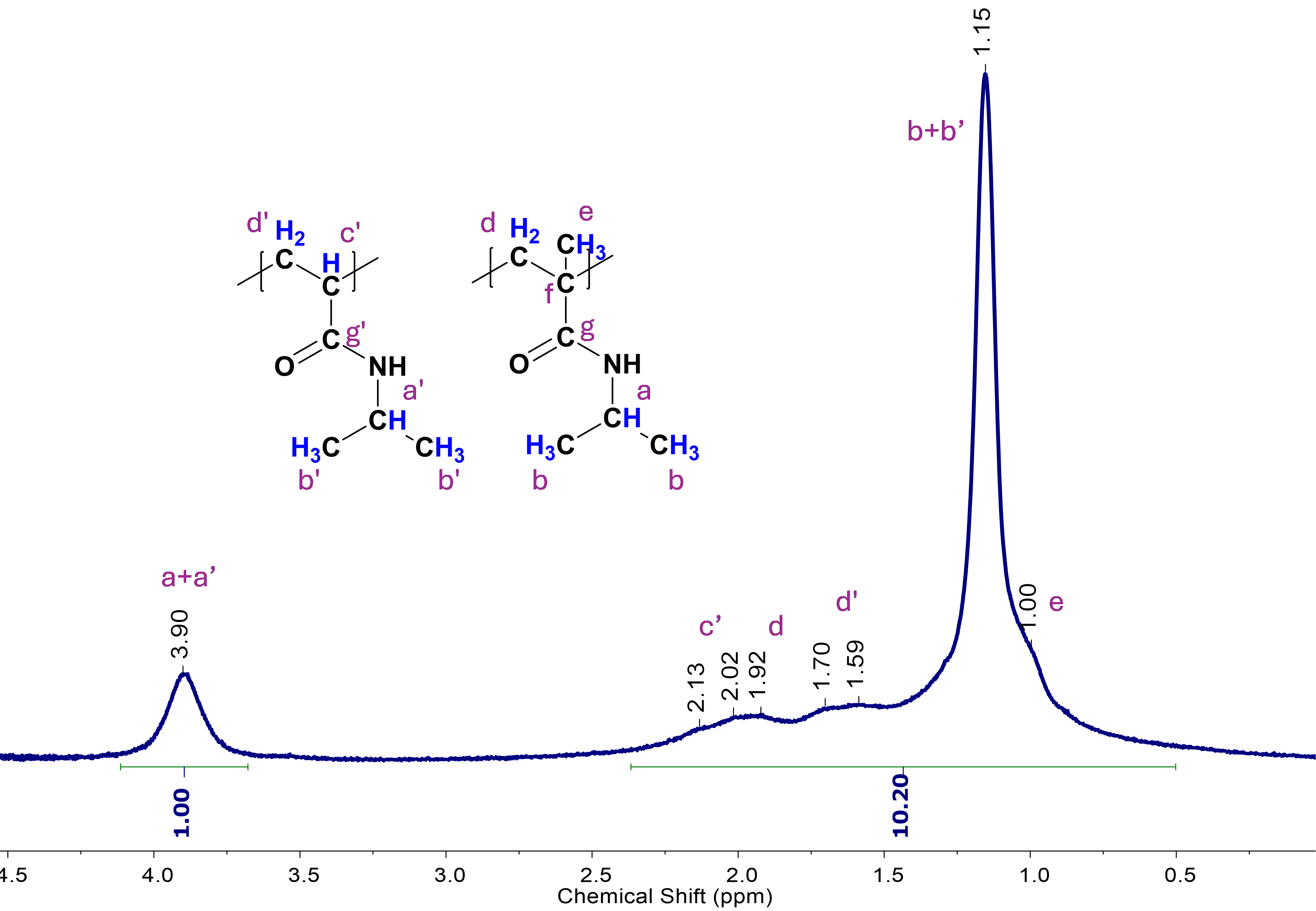}
\caption{$^1$H-NMR spectrum at 25°C of P(H-NIPAM-\textit{co}-H-NIPMAM) microgel dispersion in D$_2$O.}
\label{fig:SI_NMR_3}
\end{figure*}

\begin{figure*}
\centering
\includegraphics[width=0.8\linewidth]{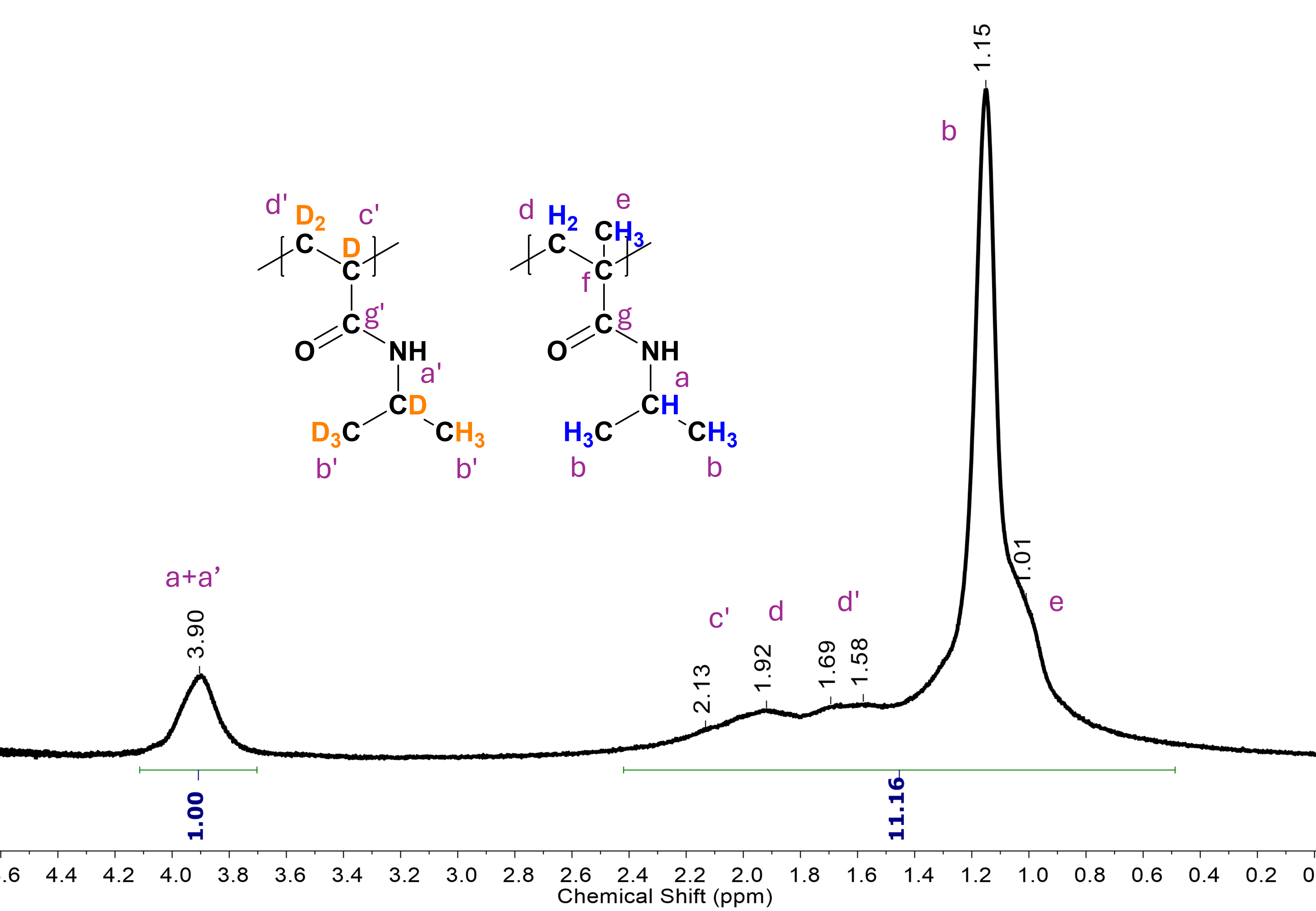}
\caption{$^1$H-NMR spectrum at 25°C of P(D-NIPAM-\textit{co}-H-NIPMAM) microgel dispersion in D$_2$O. The D-NIPAM is a mixture of 32\% mol of NIPAM and 68\% mol of NIPAM-d$_{10}$}
\label{fig:SI_NMR_4}
\end{figure*}

\begin{figure*}
\centering
\includegraphics[width=0.8\linewidth]{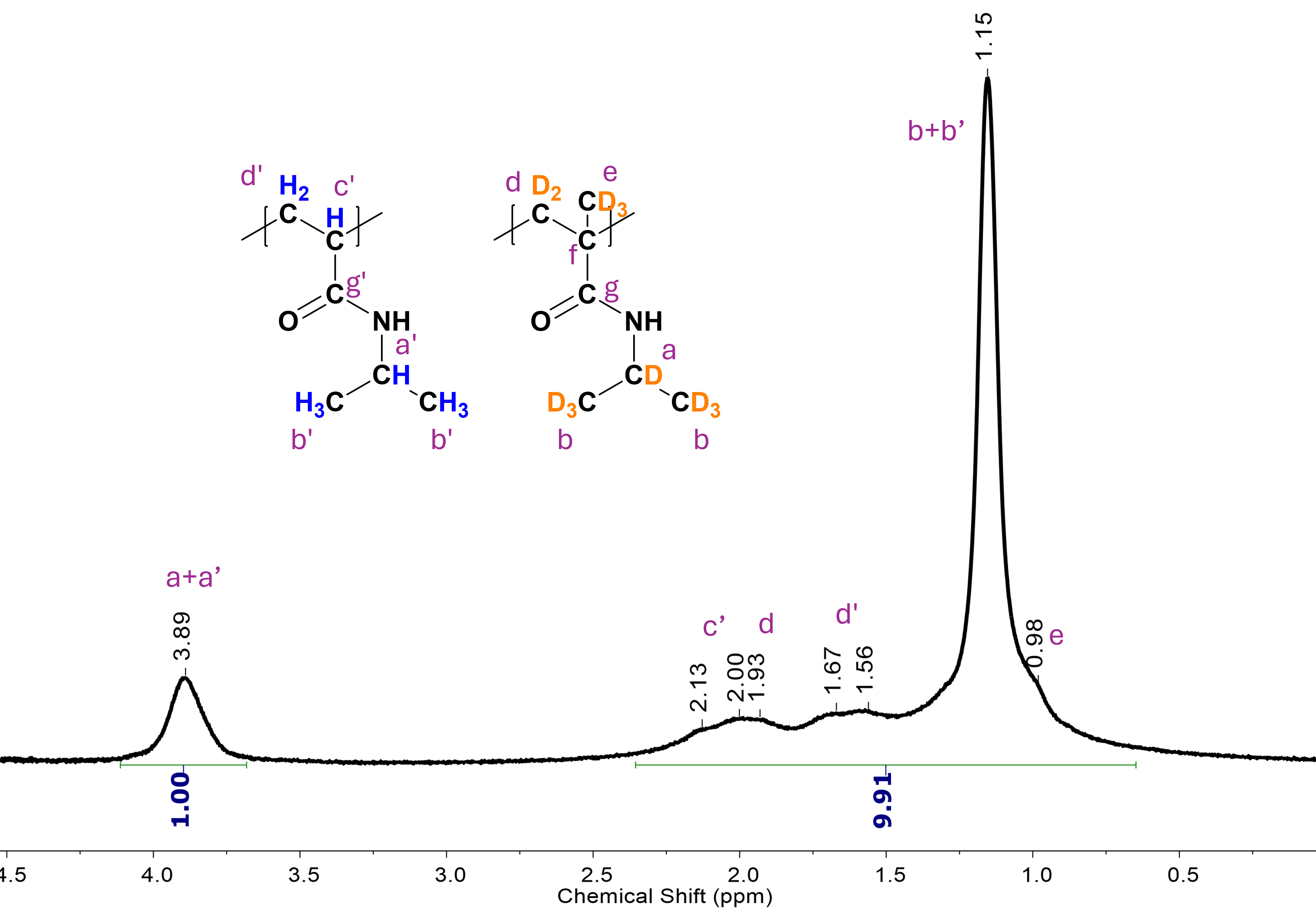}
\caption{$^1$H-NMR spectrum at 25°C of P(H-NIPAM-\textit{co}-D-NIPMAM) microgel dispersion in D$_2$O. The D-NIPMAM is a mixture of 44\% mol of NIPMAM and 56\% mol di NIPMAM-d$_{12}$}
\label{fig:SI_NMR_5}
\end{figure*}

To evaluate the copolymer microgels composition, we have employed $^1$H-NMR experiments. Figures below (Figures~\ref{fig:SI_NMR_1}-~\ref{fig:SI_NMR_5}) show the peaks assignment and the integration in the range between 0.5 and 2.5 ppm, used to calculate the molar composition of the copolymers.

\clearpage
\newpage
\subsection{$^{13}$C-NMR of microgels}

\begin{figure*}
\centering
\includegraphics[width=0.8\linewidth]{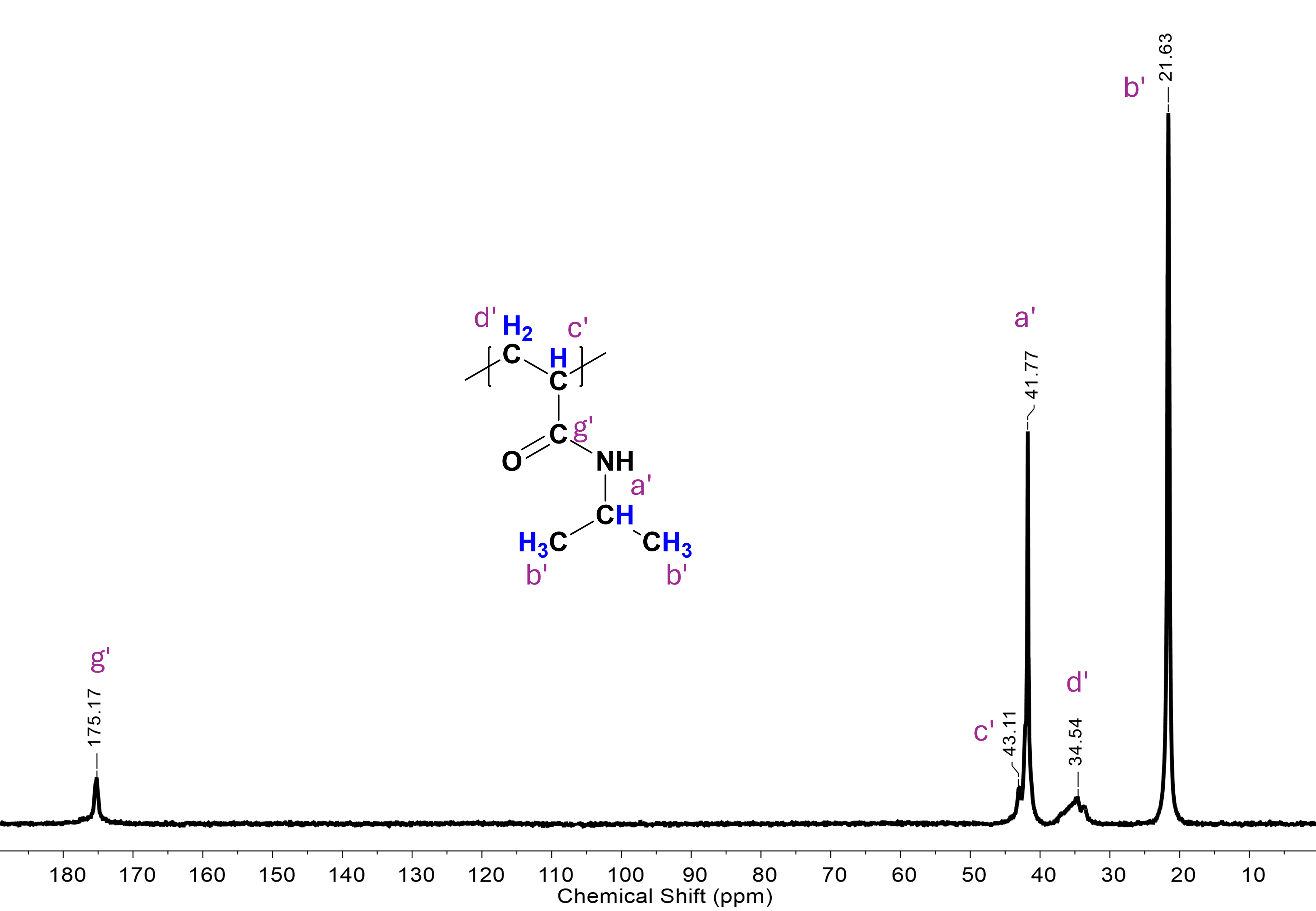}
\caption{$^{13}$C-NMR spectrum at 25°C of PNIPAM microgel dispersion in D$_2$O.}
\label{fig:SI_NMR_6}
\end{figure*}

\begin{figure*}
\centering
\includegraphics[width=0.8\linewidth]{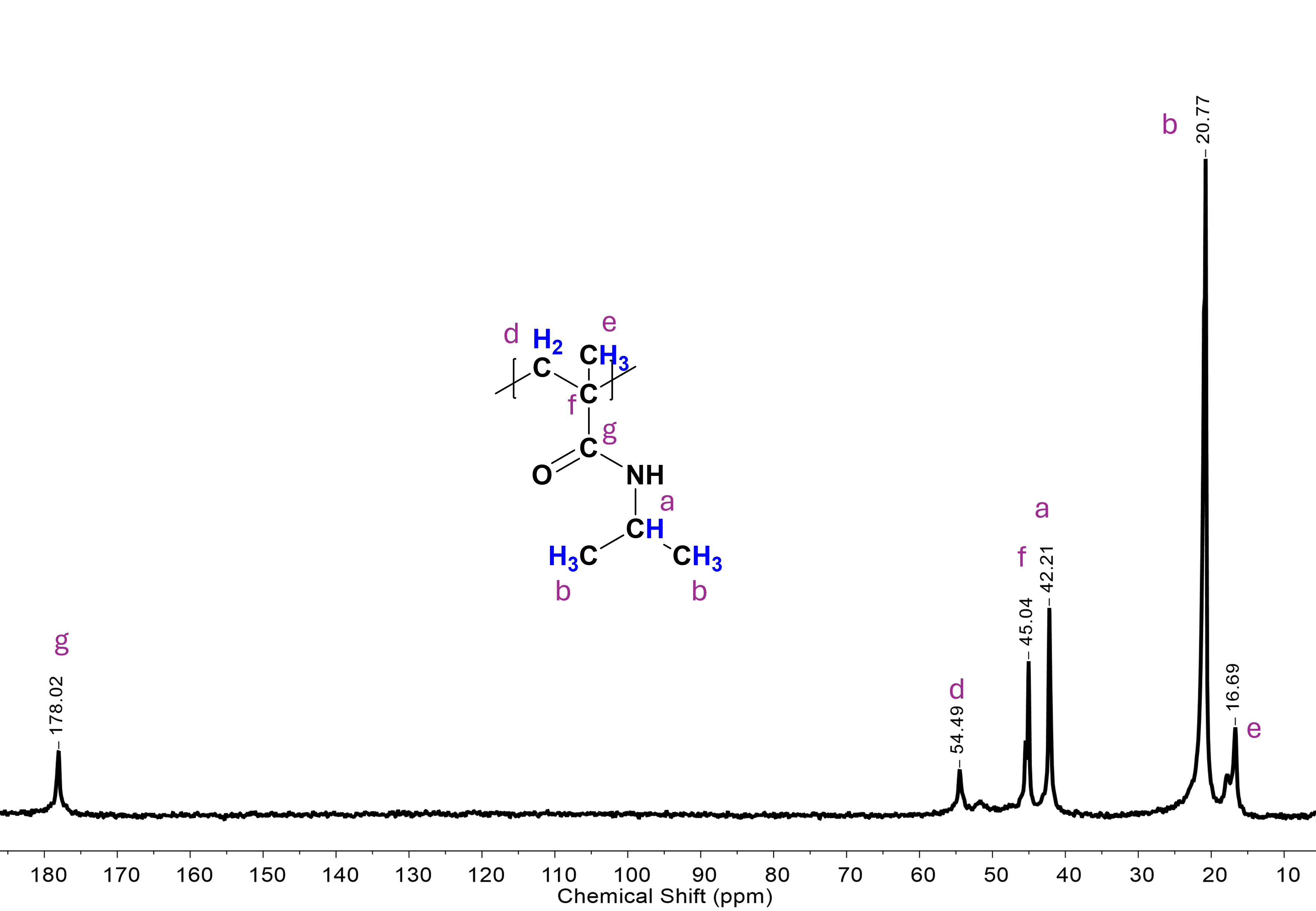}
\caption{$^{13}$C-NMR spectrum at 25°C of PNIPMAM microgel dispersion in D$_2$O.}
\label{fig:SI_NMR_7}
\end{figure*}

\begin{figure*}
\centering
\includegraphics[width=0.8\linewidth]{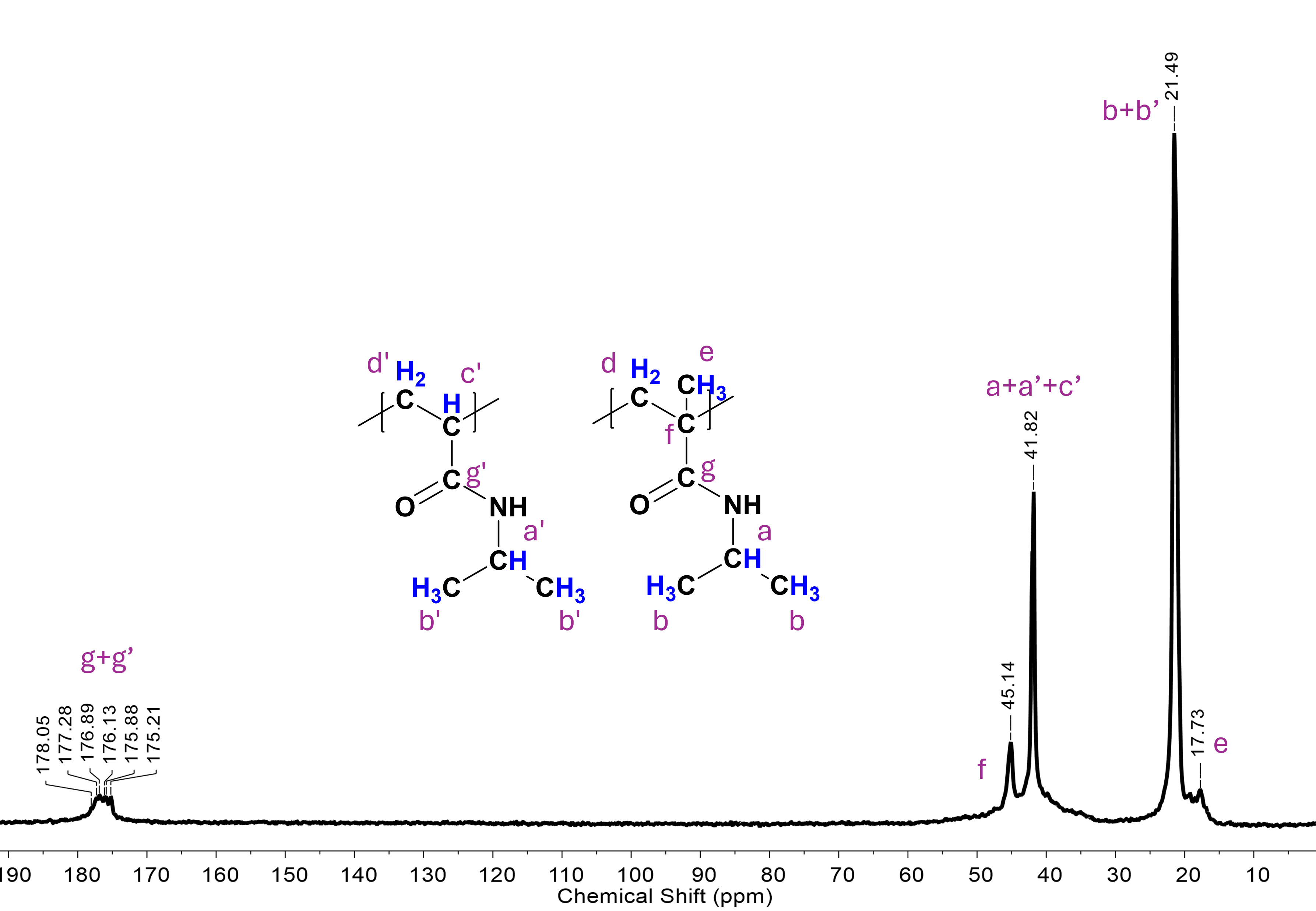}
\caption{$^{13}$C-NMR spectrum at 25°C of P(H-NIPAM-\textit{co}-H-NIPMAM) microgel dispersion in D$_2$O.}
\label{fig:SI_NMR_8}
\end{figure*}

\begin{figure*}
\centering
\includegraphics[width=0.8\linewidth]{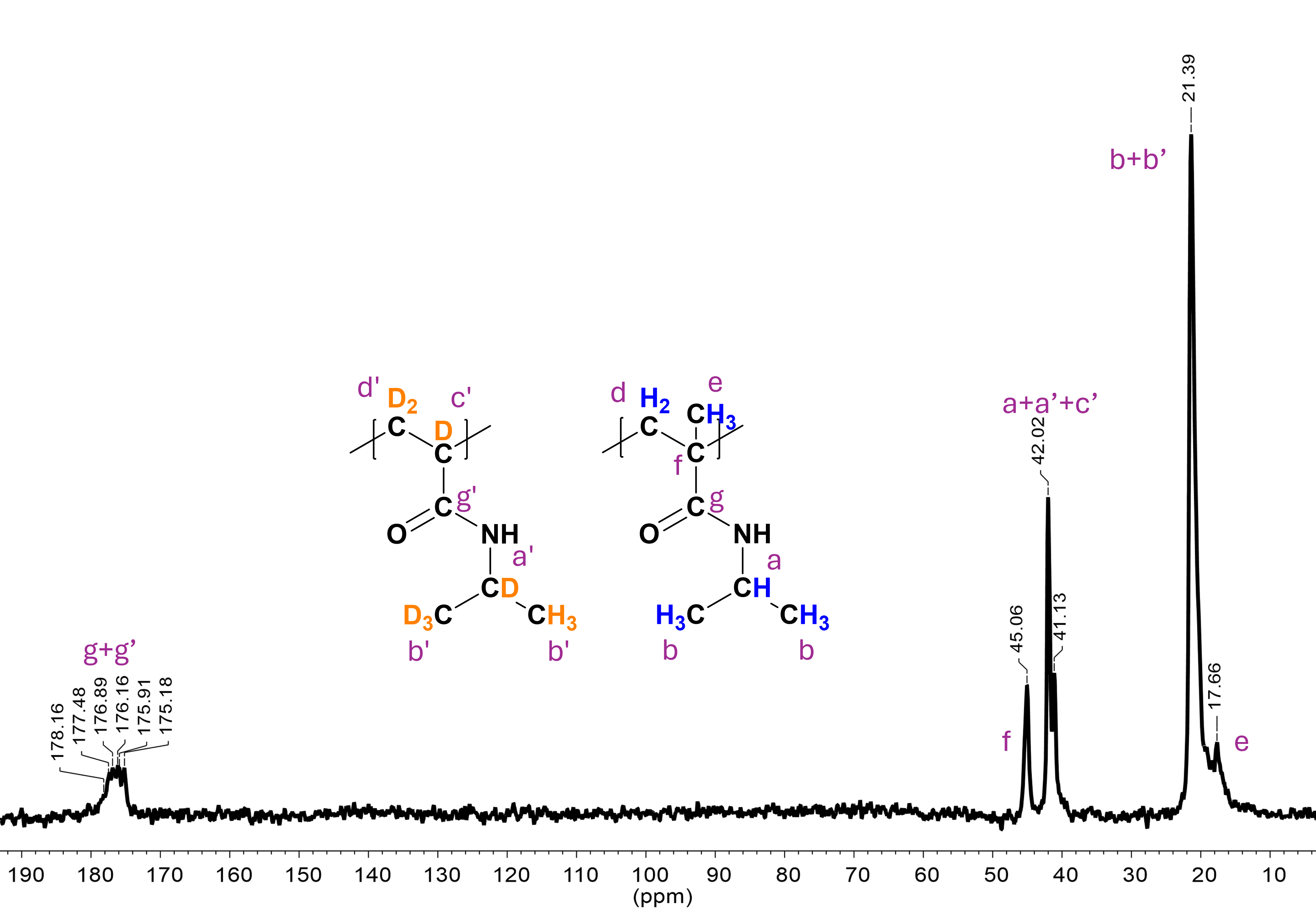}
\caption{$^{13}$C-NMR spectrum at 25°C of P(D-NIPAM-\textit{co}-H-NIPMAM) microgel dispersion in D$_2$O. The D-NIPAM is a mixture of 32\% mol of NIPAM and 68\% mol of NIPAM-d$_{10}$}
\label{fig:SI_NMR_9}
\end{figure*}

\begin{figure*}
\centering
\includegraphics[width=0.8\linewidth]{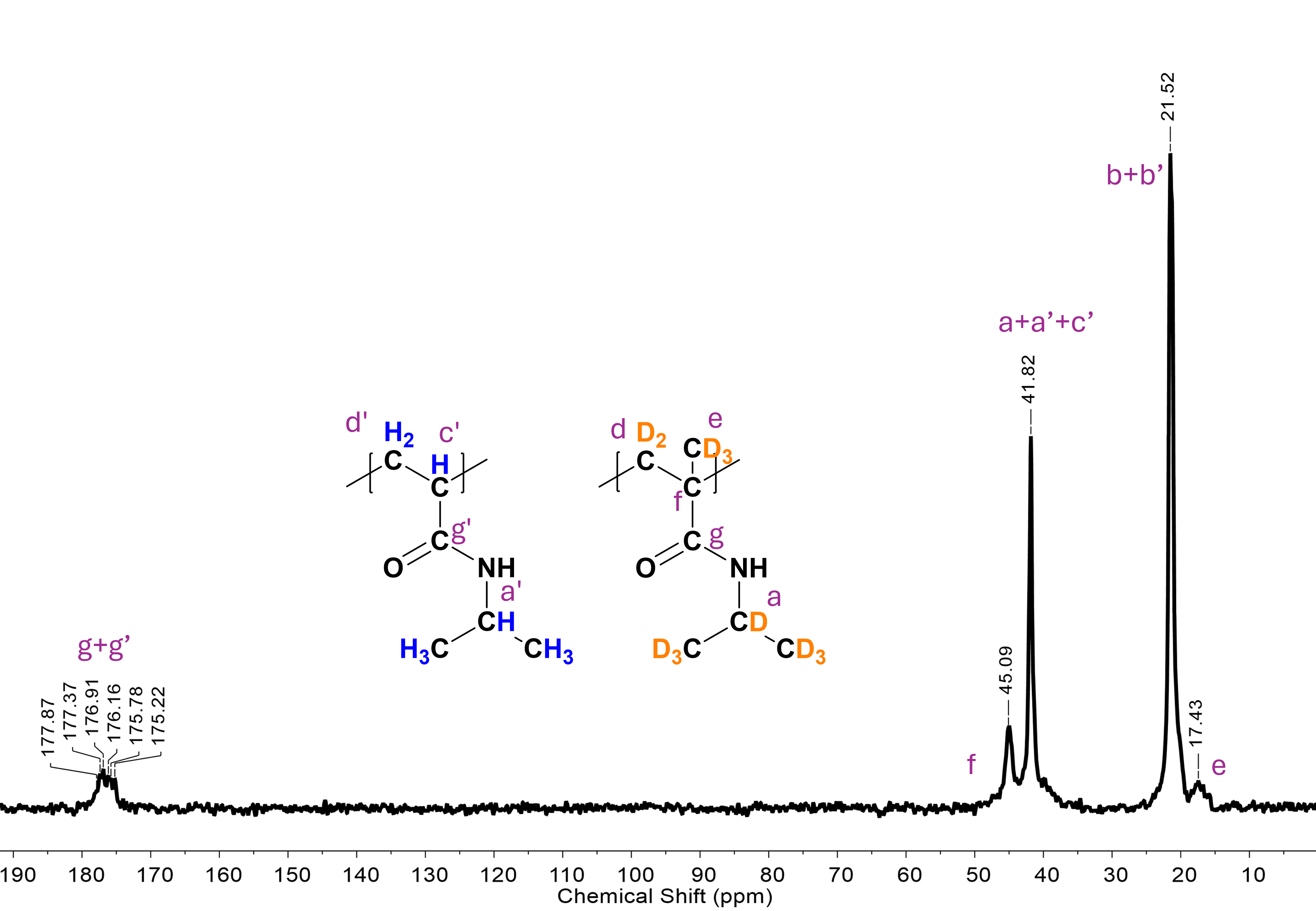}
\caption{$^{13}$C-NMR spectrum at 25°C of P(H-NIPAM-\textit{co}-D-NIPMAM) microgel dispersion in D$_2$O. The D-NIPMAM is a mixture of 44\% mol of NIPMAM and 56\% mol di NIPMAM-d$_{12}$}
\label{fig:SI_NMR_10}
\end{figure*}

\begin{figure*}
\centering
\includegraphics[width=0.8\linewidth]{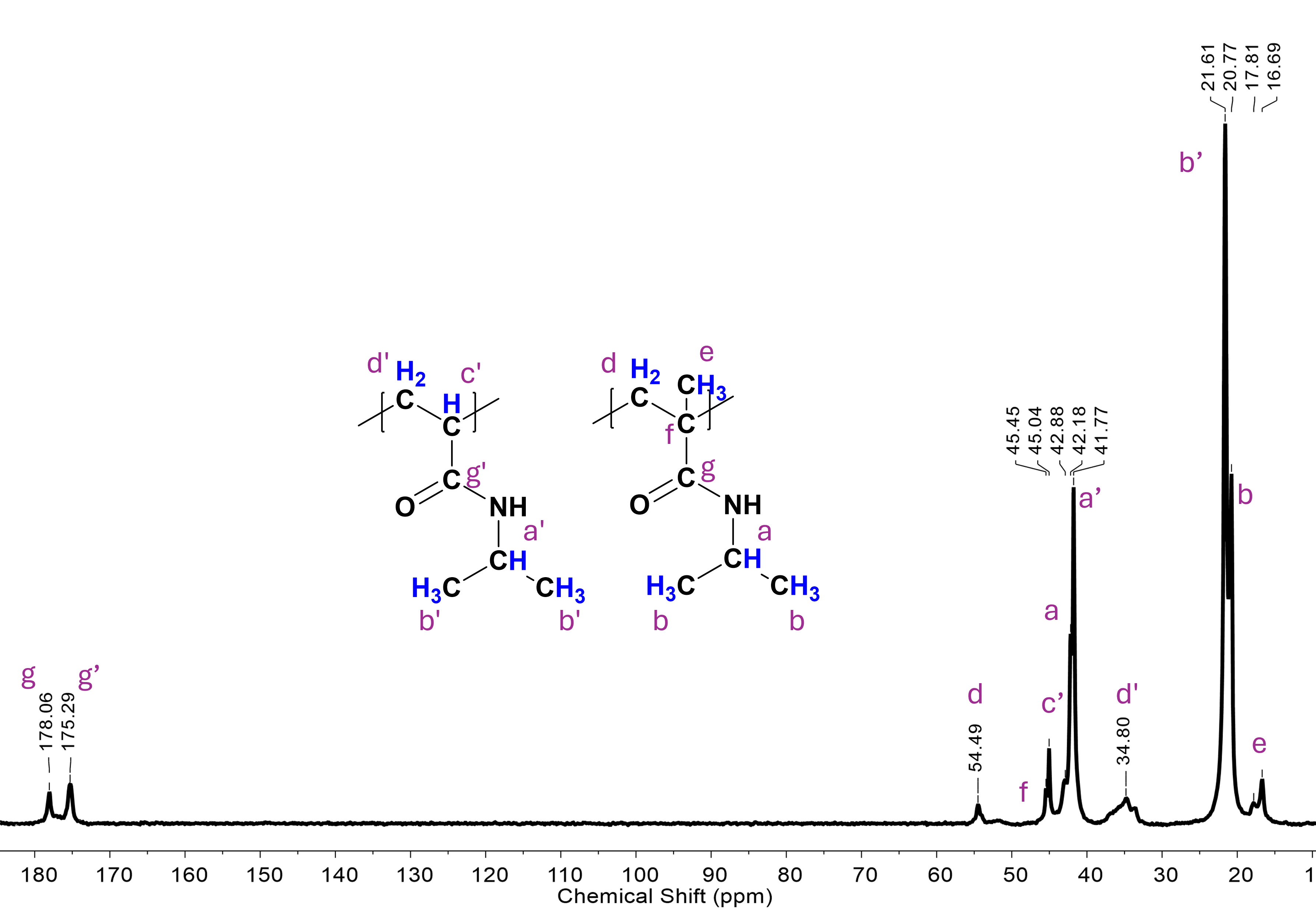}
\caption{$^{13}$C-NMR spectrum at 25°C of mixture (1/1 by weight) of PNIPAM and PNIPMAM microgels}
\label{fig:SI_NMR_11}
\end{figure*}

\clearpage
\newpage
\subsection{Carbonyl signal deconvolution in the $^{13}$C-NMR spectra}
\begin{figure*}
\centering
\includegraphics[width=0.8\linewidth]{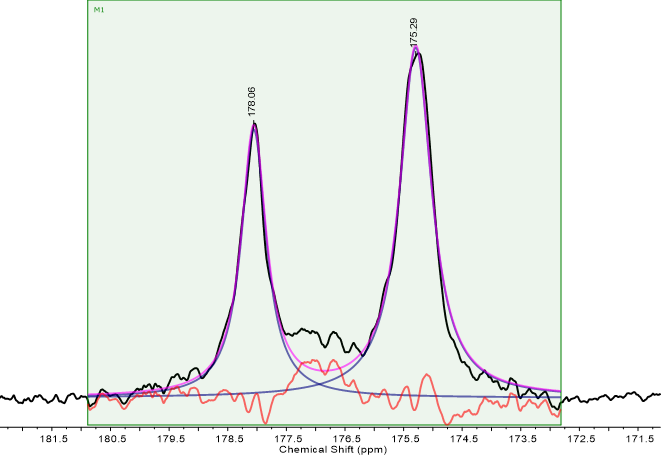}
\caption{Deconvolution of the carboxylic peaks in the $^{13}$C NMR spectrum of mixture of PNIPAM and PNIPMAM microgels}
\label{fig:SI_NMR_12}
\end{figure*}

\begin{figure*}
\centering
\includegraphics[width=0.8\linewidth]{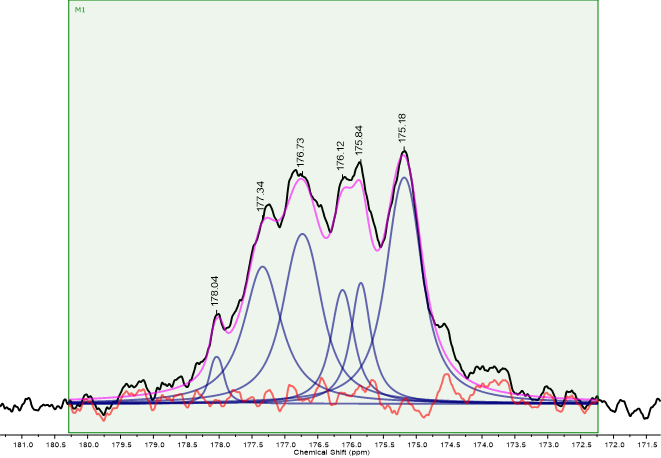}
\caption{Deconvolution of the carboxylic peaks in the $^{13}$C NMR spectrum of  P(H-NIPAM-\textit{co}-H-NIPMAM) microgel}
\label{fig:SI_NMR_13}
\end{figure*}

\begin{figure*}
\centering
\includegraphics[width=0.8\linewidth]{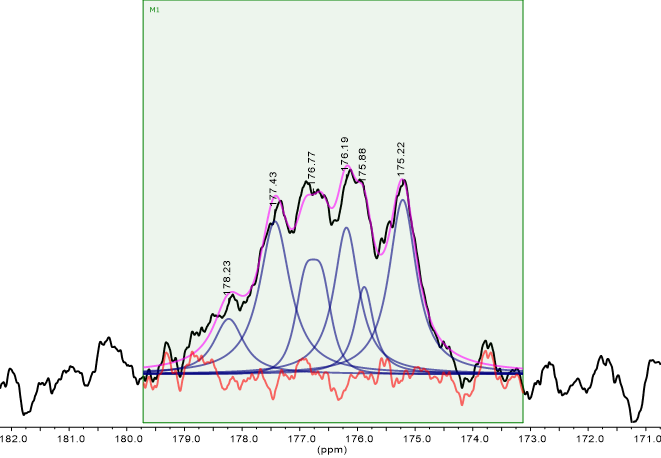}
\caption{Deconvolution of the carboxylic peaks in the $^{13}$C NMR spectrum of P(D-NIPAM-\textit{co}-H-NIPMAM) microgel}
\label{fig:SI_NMR_14}
\end{figure*}

\begin{figure*}
\centering
\includegraphics[width=0.8\linewidth]{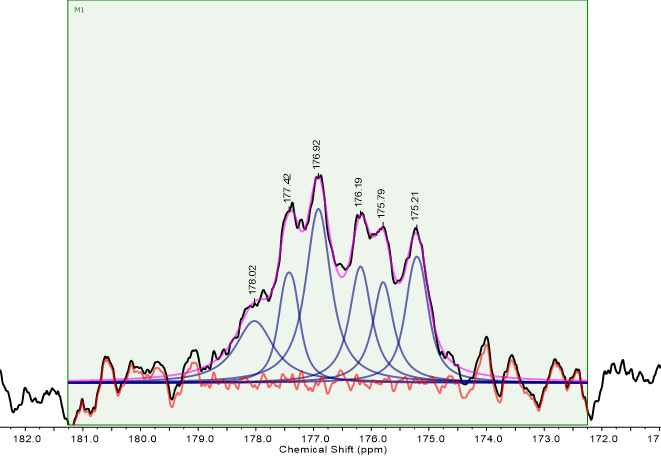}
\caption{Deconvolution of the carboxylic peaks in the $^{13}$C NMR spectrum of  P(H-NIPAM-\textit{co}-D-NIPMAM) microgel}
\label{fig:SI_NMR_15}
\end{figure*}

Deconvolution of the carbonyl signal was used to determine the relative percentage contribution of the different sequences (PPP, PPM, MPM, MMM, MMP, and PMP) to the overall peak.
In the images shown below (Figure ~\ref{fig:SI_NMR_12}-~\ref{fig:SI_NMR_15}), the black line represents the original acquired spectrum, the blue lines are the deconvoluted single peaks, the violet line is the sum of the deconvoluted peaks, almost superimposed with the original spectrum, and the red line is the difference between the original spectrum and the deconvoluted one.

\clearpage
\newpage

\subsection{Small-angle neutron scattering fit parameters}
Small-angle neutron scattering data of all microgels systems were fitted by using a fuzzy sphere model~\cite{stieger2004small}. The fitting results well reproduce the experimental data, over the entire explored $q$-range. These results reveal that all copolymer microgels have fuzzy surfaces and dense cores. The fitted parameters are listed in Tab.~\ref{tab:fitparams}.

\begin{table}[!ht]
    \centering
    \begin{tabular}{|c|ccccc|}
         \hline
        & T ($\degree$C) &  R (nm)& $\sigma$ (nm) & $\zeta$ (nm) & poly\\
         \hline
         \multirow{4}{3em} {H-H} & 20 & 43.5 $\pm$ 2 & 9.5 $\pm$ 0.5 & 6 $\pm$ 0.4 & 0.15\\
         & 35  & 38.0 $\pm$ 1.0 & 8.0 $\pm$ 0.5 & 8.0 $\pm$ 0.5 & 0.15\\
         & 37.5  & 34.0 $\pm$ 0.8 & 6.5 $\pm$ 0.4 & 8.5 $\pm$ 1.0 & 0.14\\
         & 50 & 29.0 $\pm$ 0.6 & 0.5 $\pm$ 0.1 & 0.5 $\pm$ 0.2 & 0.14\\
         \hline
            \hline
         \multirow{4}{3em} {H-D} & 20  & 44.0 $\pm$ 1.0 & 10.0 $\pm$ 1.0 & 6.5 $\pm$ 0.5 & 0.16\\
         & 35  & 40.0 $\pm$ 1.0  & 8.5 $\pm$ 0.7 & 7.0 $\pm$ 0.6 & 0.15\\
         & 37.5  & 39.0 $\pm$ 1.2 & 8.3 $\pm$ 0.4 & 8.0 $\pm$ 0.5 & 0.15\\
         & 50 & 29.0 $\pm$ 0.6 & 2.0 $\pm$ 0.2 & 4.0 $\pm$ 0.3 & 0.14\\
         \hline
          \hline
         \multirow{4}{3em} {D-H} & 20  & 46.0 $\pm$ 2.0 & 10.0 $\pm$ 1.0 & 7.5 $\pm$ 0.8 & 0.15\\
         & 35  & 40.0 $\pm$ 1.5 & 8.5 $\pm$ 0.6 & 8.5 $\pm$ 0.5 & 0.15\\
         & 37.5  & 37.0 $\pm$ 1.5 & 7.5 $\pm$ 0.5 & 8.8 $\pm$ 0.8 & 0.14\\
         & 50 & 28 $\pm$ 1.0 & 3.5 $\pm$ 0.3 & 4.5 $\pm$ 0.5 & 0.14\\
         \hline
         \end{tabular}

    \caption{Parameters obtained by fits of the experimental SANS intensities measured for dilute microgel suspensions using Eq. 1 together with Eq. 2 for the form factor $P(q)$. The last column reports the polydispersity obtained by considering a Schulz distribution of the radius. H-H, H-D and D-H indicated P(H-NIPAM-\textit{co}-H-NIPMAM), P(H-NIPAM-\textit{co}-D-NIPMAM) and P(D-NIPAM-\textit{co}-H-NIPMAM), respectively.}
    \label{tab:fitparams}
\end{table}

\begin{figure*}
\centering
\includegraphics[width=1\linewidth]{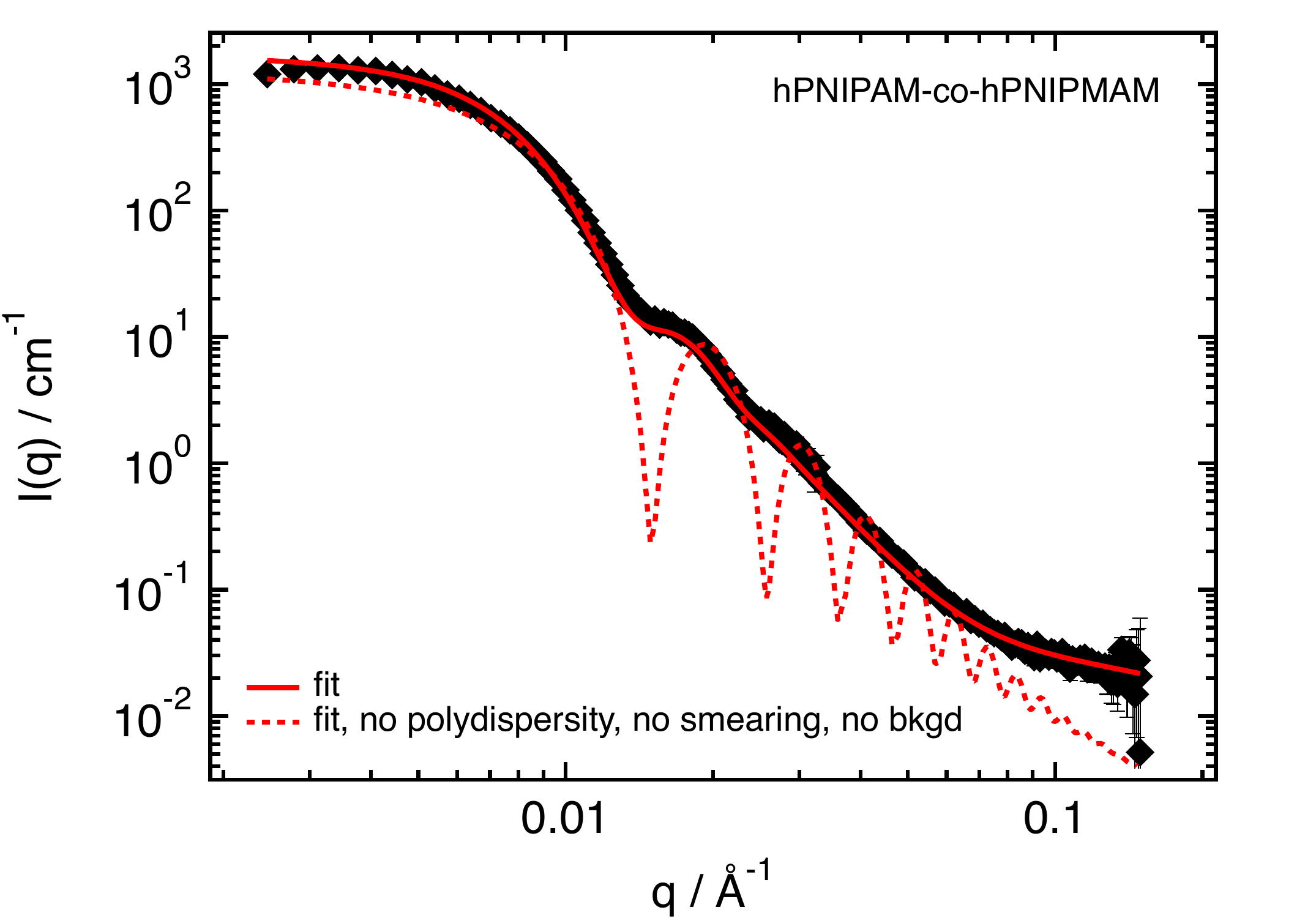}
\caption{Form factor at 20°C of P(H-NIPAM-\textit{co}-H-NIPMAM) microgels including fit with equation Eq. 2 including resolution smearing and polydispersity (solid line) and after removal of resolution smearing and polydispersity (dashed line).}
\label{fig:desmeared}
\end{figure*}

\newpage
\subsection{Small block topology of P(NIPAM-\textit{co}-NIPMAM) microgels}
\begin{figure*}
\centering
\includegraphics[width=0.3\linewidth]{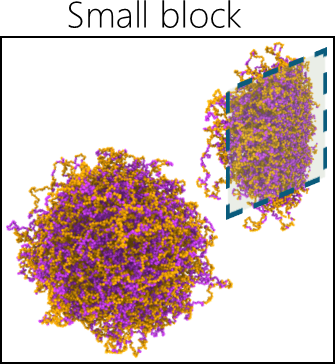}
\caption{Representative snapshot from the monomer-resolved simulations showing a small block microgel topology. PNIPAM and PNIPMAM particles are shown in purple and yellow, respectively. Cross-linkers are shown in purple. The inner part of the microgels structures is also highlighted with a sliced representation on the top panels.}
\label{fig:small}
\end{figure*}

\begin{figure*}
\centering
\includegraphics[width=1\linewidth]{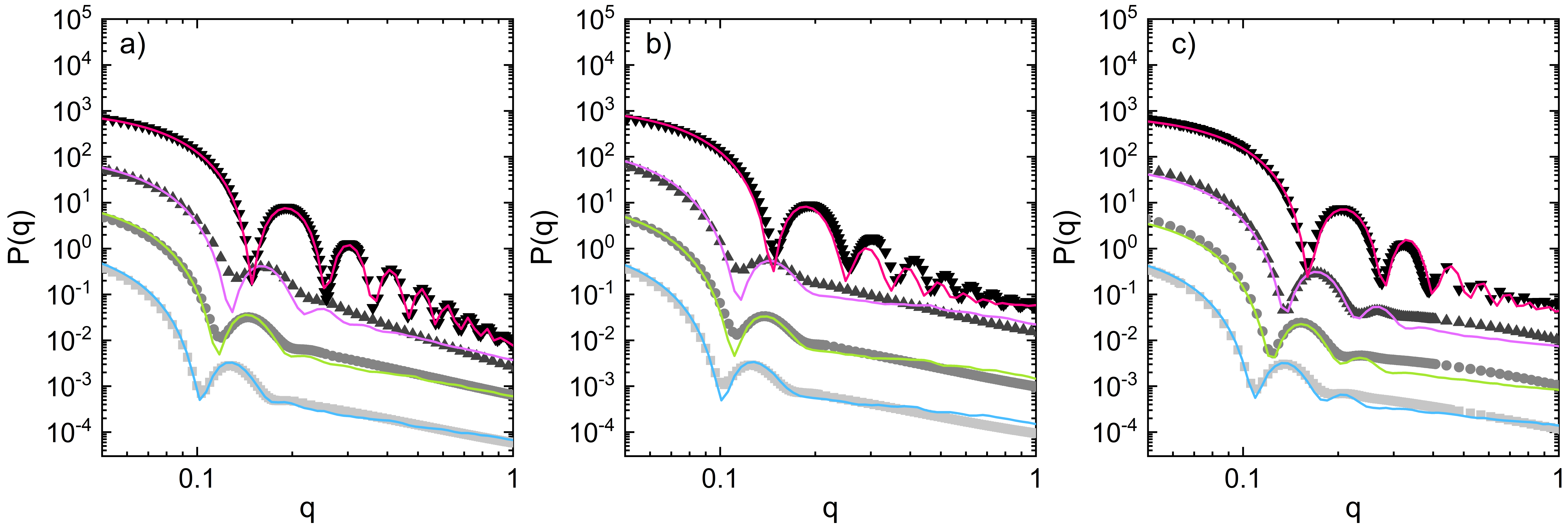}
\caption{Comparison between the experimental form factors (P(q)) measured for (a) P(H-NIPAM-\textit{co}-H-NIPMAM), (b) P(H-NIPAM-\textit{co}-D-NIPMAM), and (c) P(D-NIPAM-\textit{co}-H-NIPMAM) microgels and the numerical one calculated for the \emph{small block} topology. Experimental data are measured at 20°C (light gray squares), 35°C (gray circles), 37.5°C (dark gray triangles), and 50°C (black diamonds) and compared to numerical data calculated at the corresponding $\alpha$ values used for the other model topologies.  Data at different T values are vertically shifted for clarity.}
\label{fig:ffsmall}
\end{figure*}

The \emph{block} topology model used in this study was designed to provide a description of an extreme case of microgel internal architecture, exhibiting local domain structures. We defined domains consisting of polymer segments located between two cross-linker particles in order to generate sufficiently large regions. To further investigate the role of domain size, we developed an additional topology, referred to as \emph{small block} (Figure~\ref{fig:small}). This model was generated by randomly distributing blocks containing roughly ten consecutive PNIPMAM monomers onto a PNIPAM microgel network. The comparison between experimental and simulated form factors is shown in Figure~\ref{fig:ffsmall}. These findings reveal that the \emph{small block} topology also provides a satisfactory description of the experimental SANS data, thus supporting the hypothesis that preferential formation of local domain structures occurs within the microgel network.

\clearpage
\newpage
\subsection{Polymer-water hydrogen bonding analysis}
The occurrence of polymer-water hydrogen bond was determined by using the geometric criteria of an acceptor-donor distance lower than 0.35 nm and a hydrogen-donor-acceptor angle lower than 30$\degree$. To account for the effect of the local environment, polymer-water hydrogen bonds were calculated for each polymer repeating unit and averaged over all repeating units of the different chain models showing the same environment.

\begin{figure*}
\centering
\includegraphics[width=0.8\linewidth]{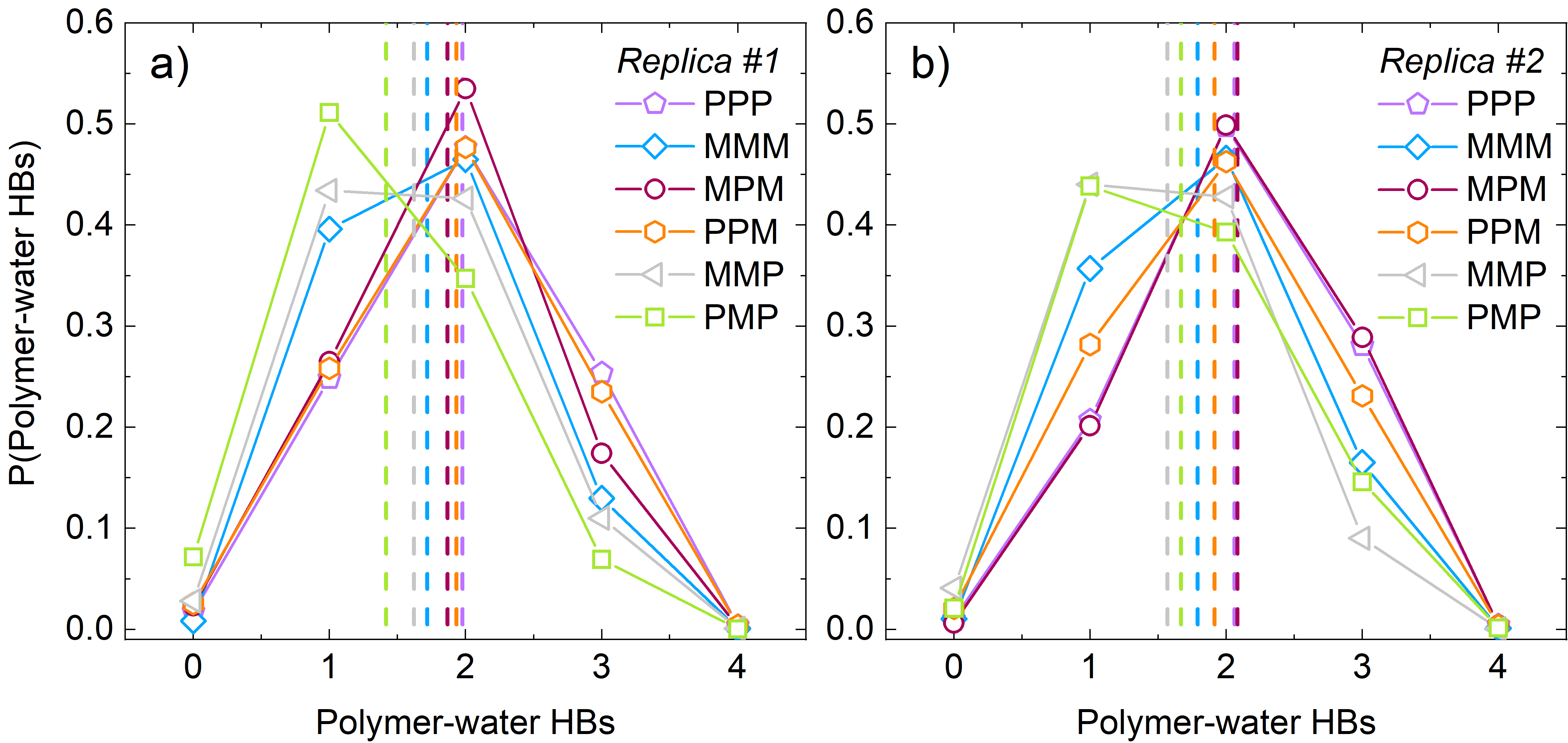}
\caption{Probability distribution of the number of polymer-water hydrogen bonds formed by each repeating unit in the atomistic chains as a function of the neighbouring repeating units: PPP (purple pentagons); MMM (blue diamonds); MPM (red circles); MPP (orange hexagons); MMP (gray triangles); and PMP (green squares). Data are calculated at 298 K and averaged over all model systems for a) the first and b) the second replica. Vertical lines are averaged distribution values.}
\label{fig:replica}
\end{figure*}

Specifically, the probability distribution of polymer–water hydrogen bonds for each NIPAM (denoted as P) and NIPMAM (denoted as M) repeating unit was classified on the basis of their neighboring repeating units in six types: (PPP) a PNIPAM unit in between two PNIPAM units; (MMM) a PNIPMAM unit in between two PNIPMAM; (MPM) a PNIPAM unit in between two PNIPMAM units; (MPP) a PNIPAM unit in between a PNIPAM and a PNIPMAM unit; (MMP) a PNIPMAM unit in between a PNIPAM and a PNIPMAM unit; and (PMP) a PNIPMAM unit in between two PNIPAM units. The PNIPAM and PNIPAMAM chains contain 88 PPP and 88 MMM configurations, respectively. The block chain consists of 43 PPP, 43 MMM, 1 MMP, and 1 MPP configurations. The random chain is composed by 7 PPP, 22 PPM, 14 PMP, 6 MMM, 24 MMP, and 15 MPM configurations. The analysis was carried out over the last 500 ns of trajectory data, sampled every 100 ps. The probability distribution of the number of polymer-water hydrogen bonds was calculated for two independent replicas, as shown in Figures~\ref{fig:replica}a and~\ref{fig:replica}b, which confirm the robustness of results.

\subsection{Mapping temperature in coarse-grained simulations}
\begin{figure*}
\centering
\includegraphics[width=0.8\linewidth]{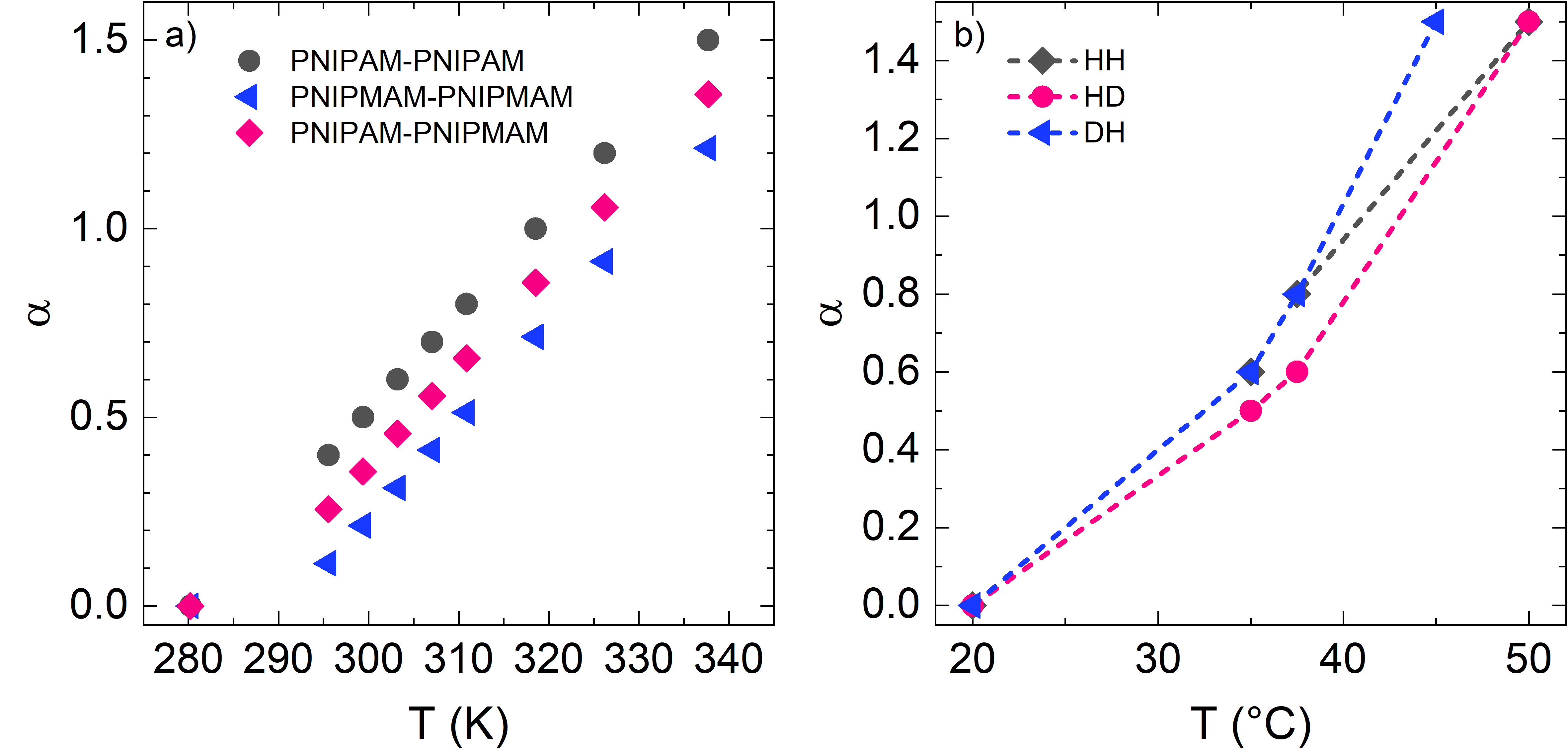}
\caption{a) Values of the solvophobicity parameter $\alpha$ employed to describe PNIPAM-PNIPAM, PNIPMAM-PNIPMAM and mixed interactions at each temperature. b) $\alpha$-temperature relation for H-PNIPAM-co-H-PNIPMAM (black diamonds), H-PNIPAM-co-D-PNIPMAM (magenta circles), and D-PNIPAM-co-H-PNIPMAM (blue diamonds) microgels.}
\label{fig:alpha}
\end{figure*}

The different volume phase transition temperature of PNIPAM and PNIPMAM microgels is accounted into the \textit{in silico} model by assigning a different value of $\alpha$ to each monomer, based on the relation between $\alpha$ and temperature determined for PNIPAM microgels in Ref.~\citenum{gnan2017silico}. For mixed interactions occurring between PNIPAM and PNIPMAM monomers the average value of $\alpha$ is used. Figure~\ref{fig:alpha}a summarizes the temperature mapping used to model the monomer interactions in  this study. To compare experimental and numerical form factors and reproduce the temperature-dependent evolution of the experimental form factors, we adjust the value of the solvophobic parameter $\alpha$, keeping consistent across the different microgel topologies. The resulting mapping between $\alpha$ and temperature is reported in Figure Figure~\ref{fig:alpha}b.

\end{document}